\documentclass[aps,prb,eqsecnum,twocolumn,citeautoscript,floatfix]{revtex4}
\usepackage{amsmath,amssymb,graphicx,mathrsfs}
\usepackage{bbold}
\makeatletter

\def\ket#1{\mathinner{|{#1}\rangle}}
\def\braket#1{\mathinner{\langle{#1}\rangle}}

{\catcode`\|=\active 
  \gdef\Braket#1{\left<\mathcode`\|"8000\let|\BraVert {#1}\right>}}
\def\BraVert{\egroup\,\mid@vertical\,\bgroup}
{\catcode`\|=\active
  \gdef\set#1{\mathinner{\lbrace\,{\mathcode`\|"8000\let|\midvert #1}\,\rbrace}}
  \gdef\Set#1{\left\{\:{\mathcode`\|"8000\let|\SetVert #1}\:\right\}}}
\def\midvert{\egroup\mid\bgroup}
\def\SetVert{\egroup\;\mid@vertical\;\bgroup}
\begingroup
 \edef\@tempa{\meaning\middle}
 \edef\@tempb{\string\middle}
\expandafter \endgroup \ifx\@tempa\@tempb
 \def\mid@vertical{\middle|}
\else
 \let\mid@vertical\vrule
\fi
\makeatother

\newcommand{\cx}[1]{{c_{#1}^{\phantom{\dagger}}}}
\newcommand{\cxd}[1]{{c_{#1}^{\dagger}}}
\newcommand{\ck}[1]{{\tilde{c}_{#1}^{\phantom{\dagger}}}}

\newcommand{\mean}[1]{\ensuremath{\braket{#1}}}
\newcommand{\nbar}{\ensuremath{\bar{n}}}

\DeclareMathOperator{\im}{\mathrm{i}}

\newcommand{\one}{\ensuremath{\mathbb{1}}}

\begin{document}
\title{Operator-Based Truncation Scheme Based on the Many-Body Fermion 
Density Matrix}
\author{Siew-Ann \surname{Cheong}}
\author{Christopher L. \surname{Henley}}
\affiliation{Laboratory of Atomic and Solid State Physics, Cornell University}
\date{\today}

\begin{abstract}
\bigskip
In an earlier work [S. A. Cheong and C. L. Henley, cond-mat/0206196 (2002)],
we derived an exact formula for the many-body density matrix $\rho_B$ of a
block of $B$ sites cut out from an infinite chain of noninteracting spinless
fermions, and found that the many-particle eigenvalues and eigenstates of
$\rho_B$ can all be constructed out of the one-particle eigenvalues and
one-particle eigenstates respectively.  In this paper we improved upon this
understanding, and developed a statistical-mechanical analogy between the
density matrix eigenstates and the many-body states of a system of 
noninteracting fermions.  Each density matrix eigenstate corresponds to a 
particular set of occupation of
single-particle pseudo-energy levels, and the density matrix eigenstate
with the largest weight, having the structure of a Fermi sea ground state,
unambiguously defines pseudo-Fermi level.  Based on this analogy, we outlined 
the main ideas behind an operator-based truncation of the density matrix 
eigenstates, where single-particle pseudo-energy levels far away from
the pseudo-Fermi level are removed as degrees of freedom.  We report numerical
evidence for scaling behaviours in the single-particle pseudo-energy spectrum
for different block sizes $B$ and different filling fractions $\nbar$.  With
the aid of these scaling relations, which tells us that the block size $B$
plays the role of an inverse temperature in the statistical-mechanical
description of the density matrix eigenstates and eigenvalues, we looked into
the performance of our operator-based truncation scheme in minimizing the 
discarded density matrix weight and the error in calculating the dispersion 
relation for elementary excitations.  This performance was compared against
that of the traditional density matrix-based truncation scheme, as well as
against a operator-based plane wave truncation scheme, and found to be very 
satisfactory.
\end{abstract}

\maketitle

\section{Introduction}

An unsolved problem in many-body physics is to augment numerical exact
diagonalization, which is only feasible on comparatively small systems, so
as to give the maximum information about the thermodynamic limit.  We know
from the renormalization group that only a few degrees of freedom, \emph{viz.}
those with low energies and long wavelengths, really matter.  Hence it should
be possible in principle to discard the irrelevant parts of the Hilbert space,
but no method has been developed in real space for higher-dimensional,
interacting lattice models.  The density matrix renormalization group 
(DMRG)\cite{white92,white93,ostlund95,white98} is the only successful
method to date, but it is limited to one-dimensional chains, or 
two-dimensional systems that can be forced into one dimension as 
strips.\cite{noack94,liang94}

To get more mileage out of density matrix-based renormalization group methods,
surely we must develop a deep understanding of the structure of density
matrices of the simplest possible systems, for which analytic results are
available to guide us.  Therefore, it is appropriate to begin by considering
the ground state of a one-dimensional chain of noninteracting spinless fermions
described by a nearest-neighbour hopping Hamiltonian
\begin{equation}\label{eqn:hopping}
H = -\sum_j\left[\cxd{j}\cx{j+1} + \cxd{j+1}\cx{j}\right].
\end{equation}
Here we observe that, in one dimension, one has the same Fermi sea ground 
state for a variety of translationally-invariant Hamiltonians with hoppings
to further neighbors, provided their dispersion relation is monotonic so that
the Fermi surface occurs at the same wavevector for a given filling.
A `block' is then identified within this one-dimensional system by choosing
$B$ sites that need not be contiguous, following which we can define the
many-body density matrix $\rho_B$ of the block starting from the 
zero-temperature many-body wavefunction and tracing out all sites outside the 
chosen block of $B$ sites.  The basis of this paper is the simple factorized 
nature of $\rho_B$ and its eigenfunctions, which we derived exactly for
noninteracting (spinless or spinfull) fermions,\cite{cheong02} by extending a 
technique introduced by Chung and Peschel\cite{chung01}.

In Ref.~\onlinecite{chung01}, it was shown that many-particle density matrix
eigenstates are built from a set of single-particle creation operators and
eigenvalues, quite analogous to the energy eigenstates of a noninteracting
system of fermions.  For a block of $B$ sites cut out from a larger overall
system, there are $B$ such \emph{pseudo-creation operators} $f_l^{\dagger}$,
and the block's Hilbert space is therefore spanned by all $2^B$ products of
the $f_l^{\dagger}$'s.  The density matrix of the block can be written as
\begin{equation}\label{eqn:smform}
\rho_B = \mathscr{Q}^{-1}\exp(-\tilde{H}),
\end{equation}
which is the exponential of a \emph{pseudo-Hamiltonian} given by
\begin{equation}\label{eqn:pseudoH}
\tilde{H} = \sum_{l=1}^B \varphi_l f_l^{\dagger} f_l,
\end{equation}
with \emph{single-particle pseudo-energy spectrum} $\varphi_l$.

This paper is devoted entirely to noninteracting fermions, because the
analytic results of Ref.~\onlinecite{cheong02} permit many calculations
that could not be done by numerical brute force in an interacting case, or
can be carried out only by quite different methods (such as Monte Carlo).
Furthermore, our hypothetical renormalization or projection algorithm using
DM truncation would first be applied to a well-understood system in which
the low-energy excitations behave as noninteracting quasiparticles (as in 
a Fermi liquid or a $d$-wave superconducting phase).  Our imagined numerical
method in such a system would extract renormalized pseudo-creation operators
$\tilde{f}_l^{\dagger}$ which are related to the single-particle operators 
$\{f_l^{\dagger}\}$ in the same way that quasiparticle creation operators 
are related to bare fermion creation operators.  The pseudo-dispersion 
relation for the renormalized $\{\tilde{f}_l^{\dagger}\}$ is thus expected 
to scale in the same fashion as the pseudo-dispersion $\varphi_l$ encoded in
\eqref{eqn:pseudoH} of the noninteracting fermions considered in this paper.

Through the calculations in this paper we aim to understand the analytic
structure of $\rho_B$, and to begin to understand the quantitative errors
due to truncation.  This includes the question: what is the proper measure of
error?  The most familiar measure, the retained fraction of total density
matrix weight, does not seem to be the best measure of error, as evidenced
by the small errors obtained in Section~\ref{sect:cdmts} for the calculations
of the dispersion relation.  Another question to be investigated is whether
--- when we are severely truncating the Hilbert space, and attempting only
to obtain the low-energy excitations --- the density matrix eigenstates are
the optimal basis.  As shown by the comparison in Section~\ref{sect:cpwts},
they are certainly superior to the other plausible basis (plane waves).

In Section~\ref{sect:dmstructure}, we summarize first our much
improved understanding of the analytic structure of the density matrix
for noninteracting fermions (and by implication for any Fermi liquid)
following from \eqref{eqn:smform} and \eqref{eqn:pseudoH}, and elucidate
the statistical-mechanical analogy between the density matrix eigenstates
and the many-body states of a system of noninteracting spinless fermions.
In Section~\ref{sect:oldresults}, we summarize the important results which 
we obtained in Ref.~\onlinecite{cheong02}, giving exact relationships
between the block density matrix $\rho_B$, the Green function matrix $G$
restricted to the block, as well as their eigenstates and eigenvalues.
Then in Section~\ref{sect:recipe}, we develop the main ideas behind a 
\emph{operator-based density matrix (DM) truncation scheme} based on the 
statistical-mechanical analogy described in Section~\ref{sect:dmstructure}.

The relation between the single-particle density matrix eigenstates and
single-particle energy eigenstates of a system of noninteracting spinless
fermions also suggests how the distribution of single-particle pseudo-energies 
$\varphi_l$ are expected to scale with the block size $B$.  Numerically, 
a scaling relationship between $\varphi_l$ and $B$ was found indeed, for 
the overall chain at various fillings $\nbar$.  Our analytical 
results from Ref.~\onlinecite{cheong02} shed light on this eigenvalue scaling 
in two ways.  Firstly, as in Ref.~\onlinecite{chung01}, they allow numerical 
study of the density matrix for system sizes so large that they would be 
inaccessible to any other techniques.  Secondly, the exact connection of
the block density matrix $\rho_B$ to the block Green function matrix $G$ gives 
hints about the eigenvalue distribution.  All of these will be discussed 
further in Section~\ref{sect:scaling}, and the implications of the scaling
behaviour of the single-particle pseudo-energies are discussed in 
Section~\ref{sect:implications}, where we derived the asymptotic behaviour,
in the limit of infinite block sizes, of the largest density matrix weight
and the truncated weight $W_t$, which is the sum of weights of the density 
matrix eigenstates retained in the operator-based DM truncation scheme.

Compared to the traditional density matrix truncation scheme used in
the DMRG, our operator-based density matrix truncation scheme gives for the
same number of density matrix eigenstates retained a slightly larger
discarded weight $\epsilon = 1 - W_t$ (see FIG.~\ref{fig:neps} in 
Section~\ref{sect:residue}).  This quantity gives a $O(\epsilon)$ 
estimate as an upper bound --- a worst case scenario --- for the
error incurred when computing the expectation of a most general observable.
As with examples in numerical integration, the performance of an algorithm in 
integrating some classes of functions may be much better than that expected
from the straightforward error analysis.  In any case, we are not really 
interested in arbitrary observables, but rather, in the dispersion relation of
elementary excitations, which we calculate in Section~\ref{sect:cdmts}.
The results are highly encouraging: the dispersion relation calculated in
the operator-based DM truncation scheme differ from the true dispersion 
relation by an amount much smaller than what is suggested by the discarded 
weight.

Besides quasiparticle dispersion relations, real space correlation functions
are also interesting quantities to calculate, and these invariably depend
on the real space structure of the many-body ground state wavefunction. 
Since this wavefunction is to be written in terms of the one-particle
density matrix eigenfunctions, it is important to understand the real
space structure of these as well.  The one-particle density matrix 
eigenfunctions kept in our operator-based DM truncation scheme have spatial 
structures that are very similar to each other.  In 
Section~\ref{sect:wavefunctionscaling} we look into a representative 
one-particle density matrix eigenfunction, the pseudo-Fermi eigenfunction,
for each $B$, and found that they also obey a universal scaling relation.  
Then in Section~\ref{sect:dm2pw}, we check how well such a truncated basis of
one-particle density matrix eigenfunctions can approximate the true 
single-particle wavefunction at the Fermi level, which is a plane wave with
wavevector $k_F$.  We find the approximation to be good even when less than
one quarter of the one-particle density matrix eigenfunctions are kept.  This
is impressive, considering the fact that in the operator-based DM truncation 
scheme, the number of many-particle density matrix eigenstates thus 
represented by the one-particle density matrix eigenstates retained constitutes
a miniscule fraction of the total number of density matrix eigenstates.

Finally, for systems where we know that the true single-particle wavefunctions 
are plane waves, it seems \emph{a priori} plausible that a plane wave-based 
operator-based truncation scheme might outperform the operator-based density 
matrix truncation scheme.  We look into this possibility in 
Section~\ref{sect:cpwts}, and find that while there are a few aspects in 
calculating the dispersion relation where the \emph{operator-based plane wave 
(PW) truncation scheme} outperforms the operator-based DM truncation scheme, 
the overall performance of the PW scheme is inferior to the DM scheme.  We
then conclude in Section~\ref{sect:conclusions} by summarizing the important
results obtained in this paper, and discuss where all these might fit into
the numerical analysis of an interacting system.

\section{Operator-Based Density Matrix Truncation}
\label{sect:consistenttruncation}

\subsection{Structure of Density Matrix Eigenvalues and Eigenstates}
\label{sect:dmstructure}

Because the Hamiltonian in \eqref{eqn:hopping} conserves particle number,
the eigenstates of $\rho_B$ have definite particle number, and may be
grouped into various $P$-particle sectors, where $P = 0, 1, \dots, B$.
A consequence of our fundamental formulas \eqref{eqn:smform} and
\eqref{eqn:pseudoH} is that every eigenstate of $\rho_B$ has the form
\begin{equation}\label{eqn:dmeigenstate}
\ket{\chi_L} = \prod_{p=1}^P f_{l_p}^{\dagger}\ket{0}.
\end{equation}
Each eigenstate is specified by a list of \emph{pseudo-occupation numbers}
$\{n_l\}$, where $n_{l_p} = 1$ for the factors contained in 
\eqref{eqn:dmeigenstate}, and is zero otherwise.  Furthermore, the
corresponding eigenvalue, the \emph{density matrix weight}, is simply given by
\begin{equation}\label{eqn:dmweight}
w_L = \mathscr{Q}^{-1}\exp\left[-\sum_{l=1}^B n_l\varphi_l\right],
\end{equation}
where the quantity
\begin{equation}\label{eqn:totalpseudoenergy}
\Phi = \sum_l n_l\varphi_l
\end{equation}
appearing in the exponent is the \emph{total pseudo-energy}.  In terms of
the single-particle pseudo-energies $\{\varphi_l\}$, the normalization
constant of the density matrix in \eqref{eqn:smform} can be written as
\begin{equation}\label{eqn:grandpartitionfunction}
\mathscr{Q} \equiv \sum_{\{n_l\}}\exp\left[-\sum_l n_l\varphi_l\right],
\end{equation}
where the summation is over all $2^B$ combinations of occupations.

It is immediately clear from \eqref{eqn:dmweight} that the density matrix
eigenstate of maximum weight corresponds to the minimum total pseudo-energy.
This is obtained by setting $n_l = 1$ for $\varphi_l < 0$ and $n_l = 0$
for $\varphi_l > 0$.  In complete analogy to the real energy of a 
noninteracting system of fermions, we simply fill up the \emph{single-particle
pseudo-energy levels (PELs)} from the lowest up to a \emph{pseudo-Fermi level}.
The maximum density matrix weight always turns out to occur with the
block fractional filling that is closest to the bulk filling of the Fermi
sea ground state.  More generally, the maximum-weight state in the $P$-particle
sector is obtained by filling the states with the $P$ lowest single-particle
pseudo-energies.  Finally, it is clear that the next-highest weights,
or equivalently the next-lowest total pseudo-energies in the $P$-particle
sector, are obtained by making particle-hole excitations involving the PELs
near to the last one filled.

The above analogy may be extended to note that \eqref{eqn:smform} is
exactly the density matrix that would be obtained (see for example,
Ref.~\onlinecite{pathria96}) at temperature $T = 1$ if $\tilde{H}$ were the 
Hamiltonian.  The reciprocal of the normalization constant $\mathscr{Q}$ of 
$\rho_B$ in \eqref{eqn:grandpartitionfunction} just corresponds to
the grand partition function for the block of $B$ sites.  Among other things,
this implies that $\braket{n_l}$, the average particle number in a particular
PEL, has the functional form of the Fermi-Dirac distribution.

We will actually apply this idea in a slightly different way, so as to
relate the single-particle pseudo-energies $\varphi_l$ for different block
sizes.  If we were dealing with an actual Hamiltonian, the dispersion 
relation would imply a density of states which would be multiplied by the
system size to obtain the actual distribution of states.  Our numerical
scaling results in Section~\ref{sect:eigscale} confirm that pseudo-energies
behave similarly to real energies.

\subsection{Relation of Pseudo-Energies to Block Green Function Matrix}
\label{sect:oldresults}

In our earlier work\cite{cheong02} we obtain an exact formula
\begin{equation}\label{eqn:exactformula}
\tilde{H} = -\sum_{ij}\left[\log_e G(\one - G)^{-1}\right]_{ij}\cxd{i}\cx{j},
\end{equation}
which, with \eqref{eqn:smform}, relates $\rho_B$ to the block Green function 
matrix $G$, whose matrix elements are $G_{ij} = \mean{\cxd{i}\cx{j}}$ with $i$ 
and $j$ restricted to sites within the block.  Clearly \eqref{eqn:exactformula}
becomes \eqref{eqn:smform} when the pseudo-Hamiltonian is diagonalized.  Also,
\eqref{eqn:exactformula} tells us that the quadratic form of $\tilde{H}$ in
\eqref{eqn:exactformula} and $G$ are simultaneously diagonalizable.  If we
denote by
\begin{equation}\label{eqn:chil}
\ket{\chi_l} = f_l^{\dagger}\ket{0} = \sum_{j=1}^B \chi_l(j) \cxd{j}\ket{0},
\end{equation}
the single-particle eigenstate of $\tilde{H}$ with eigenvalue $\varphi_l$,
then $\{\chi_l(j)\}$ is the eigenvector of $G$ with eigenvalue $\lambda_l$.
The single-particle pseudo-energies are related to the eigenvalues of $G$ by
\begin{equation}\label{eqn:pseudoenergy}
\varphi_l = -\log_e\frac{\lambda_l}{1 - \lambda_l},
\end{equation}
or equivalently,
\begin{equation}\label{eqn:pseudooccupationnumber}
\lambda_l = \frac{1}{\exp\varphi_l + 1},
\end{equation}
i.e.~the eigenvalues of $G$ are the average pseudo-occupation numbers 
$\braket{n_l}$.
Note that we sometimes write $\varphi_l \to \varphi(l, B)$ to make explicit
the dependence on block size $B$.  We will assume that $\varphi_l$ are 
ordered from the most negative to the most positive values.

Another notable result that was derived in Ref.~\onlinecite{cheong02} is that
\begin{equation}\label{eqn:QnG}
\mathscr{Q}^{-1} = \det(\one - G) = \prod_{l=1}^B (1 - \lambda_l).
\end{equation}
Along with \eqref{eqn:exactformula} and \eqref{eqn:pseudoenergy},
\eqref{eqn:QnG} comprises the final ingredients that allow numerical
computation of the density matrix even in very large blocks.  Aside from
the possibilities of truncation, \eqref{eqn:dmeigenstate} through
\eqref{eqn:QnG} have completely reduced a $2^B\times 2^B$ diagonalization
problem into a $B \times B$ problem, a computational shortcut which allows
numerical studies of large blocks.

\subsection{Recipe for Operator-Based Truncation}
\label{sect:recipe}

The analytical structure of \eqref{eqn:pseudoH} hints at the proper design
of a truncation scheme.  The retained Hilbert space of a block \emph{should
not} be the span of those density matrix eigenstates whose weight exceeds a
cutoff.  Instead, we should implement a `consistent' truncation, such that
the truncated Hilbert space consists of exactly $2^{l_{\max}}$ states, built
from all combinations of $l_{\max}$ pseudo-creation operators $f_1^{\dagger}$,
\dots, $f_{l_{\max}}^{\dagger}$, acting on a block `vacuum state' $\ket{0}_B$,
and satisfying fermion anticommutation relations.

In the traditional density matrix-based truncation scheme used in DMRG,
the recipe for truncation is to sort all density matrix weights in 
descending order, and then retain only the eigenstates associated with
the weights that exceed a certain cut off.  Let us refer to this as the
\emph{weight-ranked DM truncation scheme}.  In light of our understanding
of the structure of the many-body density matrix presented in this paper,
we can see that the weight-ranked DM truncation scheme will certainly retain
the eigenstate with maximum weight, the pseudo-Fermi sea described 
in Section~\ref{sect:dmstructure}, along with eigenstates that are `particle
excitations', `hole excitations' and `particle-hole excitations' from the
pseudo-Fermi sea.  If we arrange the entire collection of many-particle 
density matrix eigenstates into a state graph, then the state graph looks
like a $B$-dimensional hypercubic lattice near the pseudo-Fermi sea.  What 
the weight-ranked DM truncation does in this state graph picture is to remove 
nodes, and in effect cut bonds, out from this hypercubic lattice, producing 
a subgraph that is much less connected and containing tenuous links.  Because
of this, when the Hamiltonian is projected onto the weight-ranked DM truncated
basis, spurious interactions are introduced.

We can apply the pseudo-energy analogy in choosing how to truncate, given
the form of the density matrix.  It is familiar, in the truncation used in
Fourier-space-based quantum renormalization groups, to discard all 
single-particle degrees of freedom except for a shell around the Fermi surface.
In the same way, let us discard all operators $f_l$ as degrees of freedom,
except those for which $|\varphi_l|$ is less than some threshold $\varphi_*$.
For all other $f_l$, we `freeze' $n_l$ by setting $n_l$ to its ground state
value
\begin{equation}
n_l = \begin{cases}
1, \quad & \text{for $\varphi_l < -\varphi_*$}; \\
0, \quad & \text{for $\varphi_l > \varphi_*$}.
\end{cases}
\end{equation}
This choice gives the maximum density matrix weight, among the eigenstates
having any particular set of $n_l$ for the retained single-particle
pseudo-energy levels.  The spirit of this truncation scheme is similar to
that used in quantum chemistry\cite{davidson62,bender66,jensen99}, except
that the notion of a Fermi surface is more fuzzy in atoms and molecules.
The idea that truncation consists of decreasing the thickness of a shell
of wavevectors around the Fermi surface appeared in the original
renormalization group for a quantum-mechanical solid-state problem
(Anderson's poor man's RG for the Kondo problem\cite{anderson70}).  This
obvious notion --- that the action is around the Fermi surface --- 
necessarily appears in every effective form of truncation intended for
a metal (see for example, Ref.~\onlinecite{xiang96} and 
Ref.~\onlinecite{dukelsky99},
among others).  However, to our knowledge all such schemes used a basis
of plane waves or of energy eigenstates.  Our variation uses PELs in analogy 
to the use of energy eigenstates in these previous problems.  Deriving from
the density matrix, it makes sense only in procedures that involving cutting
a real-space block out of a larger system.

Within this \emph{operator-based density matrix (DM) truncation scheme},
we can define an effective Hamiltonian for the truncated Hilbert space,
just by taking the matrix elements of the true Hamiltonian between all the 
retained states.  Using an operator-based truncation, this will have a
particularly clean form: first replace each creation operator 
$\cxd{j}$ by the equivalent combination of all $\{f_l^{\dagger}\}$;
then replace $f_l^{\dagger}f_l \to 1$ for each single-particle pseudo-energy
$\varphi_l < -\varphi_*$ (these are frozen to be always occupied), and
otherwise remove all terms involving the operators that are truncated.  Thus,
if the original Hamiltonian has at most $l_{\max}$-fermion terms, the same 
will be true for the truncated Hamiltonian.  This prescription shows that
such a truncation is possible for general models, once one knows the
appropriate density matrix, but in this paper we have applied it only to
noninteracting models.

\section{Scaling Behaviour of Eigenvalue Distribution}
\label{sect:scaling}

Since the many-particle density matrix eigenvalues are built, according to
\eqref{eqn:dmweight}, from the single-particle pseudo-energies, the
latter are the focus of our numerical investigations.  Now if our entire 
system consisted of the block in a pure state at $T = 0$, then every 
eigenvalue $\lambda_l$ of the block Green function matrix $G$, being the
average pseudo-occupation number $\braket{n_l}$ of a PEL, would either be
one or zero.  At $T > 0$, $\lambda_l$ follow a Fermi-Dirac distribution.  
We will see later that cutting out a finite block from a $T = 0$ system, by 
tracing over the environment of the block, has a similar effect on the 
eigenvalues of $G$ as would taking $T > 0$ when the block is the whole system.

In a translationally invariant system with filling $\nbar$ (at $T = 0$), a
fraction $\nbar$ of the eigenvalues of $G$ are one, while the rest are zero.
Cutting out a block of length $B$ must smooth out this step (much as having
$T > 0$ makes it into a Fermi-Dirac distribution), and we expect the
transition from one to zero to occur over a fraction ${\sim}1/B$.  This guess
was inspired by the analogy of pseudo-energy $\varphi_l$ in 
\eqref{eqn:pseudoH} to the real energy, which near the Fermi level scales
linearly with wavevector ${\sim}1/B$.  This $B^{-1}$ scaling suggests the
conjecture of a scaling form for the single-particle pseudo-energy like
$\varphi_l \approx B f(l/B)$, and indeed we find below just such a scaling
form.

\subsection{Pseudo-Energies and Pseudo-Occupation Numbers}
\label{sect:eigscale}

\begin{figure}[htbp]
\centering
\includegraphics[width=0.95\linewidth,clip=true]{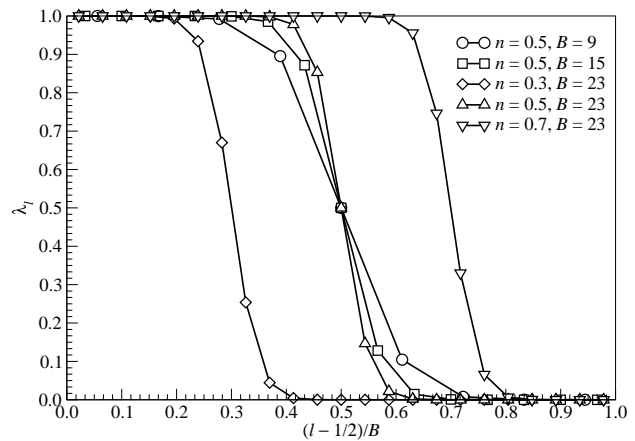}
\caption{Distribution of $\lambda_l$ for different filling fractions and 
block sizes.  In order to compare $\lambda_l$ for different block sizes, 
the interval $l \in [1, B]$ is rescaled such that $(l-\frac{1}{2})/B \in 
(0, 1)$.  With this rescaling, $\lambda = \frac{1}{2}$ always occur at
$[(l-\frac{1}{2})/B] = \nbar$.}
\label{fig:lambdadiff}
\end{figure}

In this subsection we calculate numerically the eigenvalues $\lambda_l$ of
the block Green function matrix $G$, and use \eqref{eqn:pseudoenergy} to 
compute the single-particle pseudo-energies $\varphi_l$.
For a chain of free spinless fermions in its ground state, the matrix elements 
of the block Green function matrix $G$ are
\begin{equation}
G_{ij} = \frac{\sin \pi\nbar|i-j|}{\pi |i-j|},
\end{equation}
where $\nbar$ is the filling fraction.  FIG.~\ref{fig:lambdadiff} shows how 
$\lambda_l$, the eigenvalues of $G$, are distributed for different filling 
fractions $\nbar$ and different block sizes $B$.

\begin{figure}[htbp]
\centering
\includegraphics[width=.95\linewidth,clip=true]{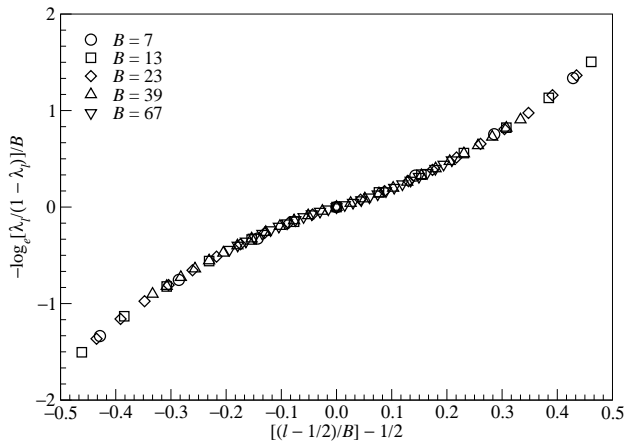}
\caption{Plot of $-B^{-1}\log_e[\lambda_l/(1 - \lambda_l)]$ as a function of 
the scaling variable $x \equiv [(l-\frac{1}{2})/B] - 1/2$ for various block 
sizes at half-filling, showing a scaling collapse onto the scaling function 
$f(x)$.  For $B > 23$, the largest and smallest pseudo-energies are severely 
affected by numerical truncation errors in the diagonalization routines, 
and thus not shown.}
\label{fig:scalingplot}
\end{figure}

For $\nbar = \frac{1}{2}$, our numerical studies suggest a scaling 
relationship of the form
\begin{equation}\label{eqn:phiscaling}
\varphi(l,B) \cong B f(x),
\end{equation}
where $x \equiv [(l-\frac{1}{2})/B] - \frac{1}{2}$, as shown in 
FIG.~\ref{fig:scalingplot}.  There are two points on 
FIG.~\ref{fig:scalingplot} we would like to note.  First of all, with our 
choice of the scaling variable $x$, the scaling function $f(x)$ always 
passes through the origin, i.e.
\begin{equation}\label{eqn:fthruorigin}
f(0) = 0.
\end{equation}
Secondly, from FIG.~\ref{fig:scalingplot}, we see that $f(x)$ is an odd
function, i.e.
\begin{equation}\label{eqn:fodd}
f(-x) = -f(x),
\end{equation}
which is what we would expect from particle-hole symmetry when the overall
system is at half-filling, and $f(x)$ has a finite positive slope at $x = 0$,
i.e.
\begin{equation}\label{eqn:fposslope}
f'(0) > 0.
\end{equation}

\begin{figure}[htbp]
\centering
\includegraphics[width=0.98\linewidth,clip=true]{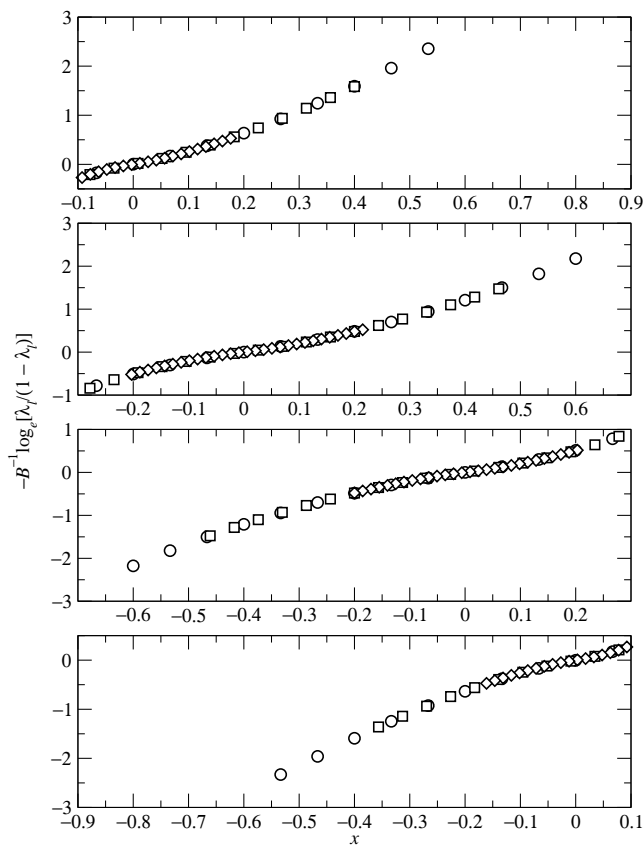}
\caption{Scaling collapses for (from top to bottom) $\nbar = 0.1$, 0.3, 0.7
and 0.9, plotted against the scaling variable $x \equiv (l-\frac{1}{2})/B -
\nbar$, for three different block sizes: $B = 15$ ($\circ$), $B = 23$ 
($\square$) and $B = 67$ ($\diamond$).}
\label{fig:completescaling}
\end{figure}

Similar scaling behaviours of the form
\begin{equation}\label{eqn:complete}
\varphi(l, B, \nbar) \cong B f(\nbar, x)
\end{equation}
are found for all $\nbar$, with the generic scaling variable 
\begin{equation}\label{eqn:scalingvariable}
x \equiv (l-l_F)/B,
\end{equation}
where
\begin{equation}
l_F = \nbar B + \tfrac{1}{2}
\end{equation}
plays the role of a Fermi wavevector, and a filling fraction-dependent scaling
function $f(\nbar, x)$, as shown in FIG.~\ref{fig:completescaling}.  The
scaling functions continue to satisfy
\begin{equation}\label{eqn:nbarorigin}
f(\nbar, 0) = 0,
\end{equation}
and
\begin{equation}
f'(\nbar, 0) > 0,
\end{equation}
but $f(\nbar, x)$ is no longer an odd function of $x$ for $\nbar \neq 
\frac{1}{2}$.  Instead, the particle-hole symmetry inherent in our model
is manifested as
\begin{equation}\label{eqn:almostodd}
f(\nbar, -x) = -f(1 - \nbar, x).
\end{equation}

From \eqref{eqn:pseudooccupationnumber} and \eqref{eqn:complete}, we can write 
$\lambda_l$ as
\begin{equation}\label{eqn:lambdascale}
\lambda_l = \frac{1}{\exp(B f) + 1},
\end{equation}
which tells us that $f(\nbar, x)$ plays the role of a dispersion relation 
$\epsilon(k)$, while $B$ plays the role of the inverse temperature $\beta$.  
This confirms our suspicion that the effect of cutting a block out of an
overall system in its ground state at $T = 0$ is to ascribe to the block
an effective temperature.  As expected, this effective temperature 
approaches zero as the block size is increased, since we are keeping more 
and more information about the overall system, which we know to be at $T = 0$.

\subsection{Normalization Constant}

Having understood that $\lambda_l$ is related to the scaling function 
$f(\nbar, x)$ as in \eqref{eqn:lambdascale}, we are now ready to investigate the
scaling behaviour of the normalization constant $\mathscr{Q}^{-1}$, which
is related to $\lambda_l$ by \eqref{eqn:QnG}.  We do
this first at half-filling.  As can be seen from FIG.~\ref{fig:lambdadiff}, 
at half-filling roughly half of the $\lambda_l$ are approximately one, whereas 
the other half are approximately zero.  The product $\prod_l (1 - \lambda_l)$ 
is therefore determined primarily by the ${\sim}B/2$ $\lambda_l$'s that are 
nearly one.  For these eigenvalues, $\exp[B f(x)] \ll 1$ and thus $1 - 
\lambda_l \approx \exp[B f(x)]$ (when it is clear what the filling fraction 
is, we will drop the argument $\nbar$ in $f(\nbar, x)$ to keep the notations
neat).  With this, we find that
\begin{equation}\label{eqn:expphi0}
\begin{split}
\mathscr{Q}^{-1} &= \prod_{l=1}^B (1 - \lambda_l) 
\approx \prod_{l=1}^{B/2} \exp[B f(x)] \\
&\approx \exp\left\{B\int_{-1/2}^0 f(x)\,dx\right\} \\
&= \exp\left\{-B\int_0^{1/2} f(x)\,dx\right\},
\end{split}
\end{equation}
where we made used the observed odd symmetry of the scaling function in
\eqref{eqn:fodd} when the overall system is at half-filling, so that the
integral within the exponent is positive.  From \eqref{eqn:expphi0}, we
see that $\mathscr{Q}^{-1}$ decreases exponentially with block size $B$.

In general, for $\nbar$ not too close to zero, where the argument that those 
$\lambda_l$'s that are near one makes the dominant contribution holds,
we find that
\begin{equation}\label{eqn:phi0scale}
\begin{split}
\mathscr{Q}^{-1} &\approx \exp\left\{B\int_{-\nbar}^0 f(\nbar, x)\,dx\right\}\\
&= \exp\left\{-B\int_0^{\nbar}f(1 - \nbar, x)\,dx\right\},
\end{split}
\end{equation}
where we made use of \eqref{eqn:almostodd}.  For $\nbar$ very close to zero,
there are a handful of $\lambda_l$'s of $O(1)$, and the rest are all nearly
zero, behaving like $\lambda_l \approx \exp[-Bf(\nbar, x)]$.  Ignoring 
these handful of $O(1)$ $\lambda_l$'s, we find that the contribution to
$\mathscr{Q}^{-1}$ from those $\lambda_l \ll 1$ is proportional to the
product $\prod_l (1 - \lambda_l) \approx 1 - \sum_l \lambda_l$, and so
\begin{equation}
\mathscr{Q}^{-1} \propto \left[1 - \int_0^{1-\nbar} 
e^{-Bf(\nbar, x)}\,dx\right].
\end{equation}
The integral can be evaluated as a cumulant expansion, but we can already see
that for large $B$, the integral will not be important, and thus 
$\mathscr{Q}^{-1}$ derives most of its value from the few $O(1)$ $\lambda_l$'s.
In contrast, when $\nbar$ is very close to 1, then most of the
eigenvalues $\lambda_l$ of $G$ are close to 1, and these continue to dominate
the product $\prod_l(1 - \lambda_l)$, and the asymptotic formula derived in
\eqref{eqn:expphi0} continues to be valid.

\section{Density Matrix Weights: Implications of Eigenvalue Scaling}
\label{sect:implications}

With our understanding of the structure of the many-particle density matrix
eigenvalues and eigenstates developed in Section~\ref{sect:dmstructure},
and on the scaling behaviour of the single-particle found in 
Section~\ref{sect:scaling}, we want to now address the question of how
much of the Hilbert space we can truncate.  Clearly, the answer to this
question depends on what measure of error we intend to use as our criteria
for judging how well the truncated Hilbert space describes the physics 
associated with the parent model.  In this section we look at the most
common measure of error, used in the DMRG\cite{white92,white93}
and quantum chemistry calculations:\cite{davidson62,bender66,jensen99} for
a properly normalized density matrix, the density matrix weights $w_L$ satisfy
the sum rule
\begin{equation}\label{eqn:sumrule}
\sum_L w_L = 1.
\end{equation}
If the ordinal numbers $L$ are chosen such that $w_L$ is ranked in decreasing
order, and a total of $L_{\max}$ density matrix eigenstates are retained,
then the truncated weight
\begin{equation}
W_t = \sum_{L \leq L_{\max}} w_L \leq 1,
\end{equation}
and the discarded weight
\begin{equation}
\epsilon = 1 - W_t
\end{equation}
are frequently used as figures of merit for the truncation scheme,
since for a bounded operator $A$, the truncation error in $\braket{A}$ is
$O(\epsilon)$.

\subsection{The Gapless Chain of Noninteracting Spinless Fermions}

\begin{figure}[htbp]
\centering
\includegraphics[width=\linewidth]{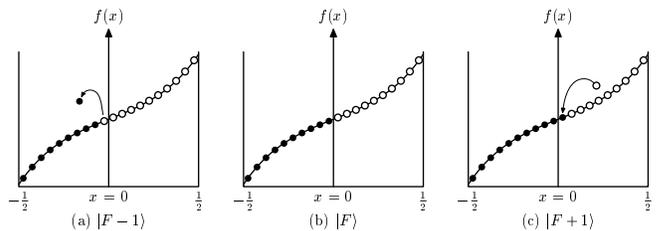}
\caption{Schematic diagram showing the three many-particle density matrix
eigenstates, (a) $\ket{F-1}$, (b) $\ket{F}$ and (c) $\ket{F+1}$, with the 
largest weights, for a block of $B$ ($B$ even) sites within a overall system
that is half-filled.}
\label{fig:fluctuation}
\end{figure}

Instead of diving in to look at $W_t$ or $\epsilon$, let us consider first a
related question: how large is the maximum weight for a block of $B$ sites
embedded within an overall system of gapless noninteracting spinless fermions?
For our discussions, we will consider the half-filled case; it will be
straightforward to extend the arguments presented below to $\nbar \neq 
\frac{1}{2}$.  For convenience, let us take $B$ to be even.\footnote{The same 
conclusion holds for the case of $B$ odd, apart from the technical annoyance 
that there are \emph{two} many-particle density matrix eigenstates with the 
largest weight.}  Let us denote by $\ket{F}$, where $F = B/2$, the 
many-particle density matrix eigenstate having the largest weight.  This
state is always kept in the operator-based truncation scheme.  Recall
from Section~\ref{sect:consistenttruncation} that, as shown in 
FIG.~\ref{fig:fluctuation}(b), this is the analog among density matrix 
eigenstates of the Fermi sea ground state among energy eigenstates.  
In this $B/2$-particle state, the single-particle pseudo-energy is filled up
to just before $x = 0$, which we shall call the \emph{pseudo-Fermi level}.
The many-particle density matrix eigenstates of the block, of which there are 
two, having the next largest weights will be called $\ket{F-1}$ 
(FIG.~\ref{fig:fluctuation}(a)) and $\ket{F+1}$ (FIG.~\ref{fig:fluctuation}%
(c)), having respectively one less or one more particle.

We can understand the weight ratio $w_{F-1}/w_{F}$ or $w_{F+1}/w_F$ as follows. 
By \eqref{eqn:dmweight}, $\ket{F}$ is the state with $n_l = 1$ for $l = 1$,
\dots, $B/2$ and $n_l = 0$ for $l = B/2 + 1$, \dots, $B$.  The state 
$\ket{F+1}$ differs only in having $n_{B/2 + 1} = 1$, while $\ket{F - 1}$
differs only in having $n_{B/2} = 0$.  Near the pseudo-Fermi level, the
scaling function has a slope of $f'(0)$, while the spacing between adjacent
pseudo-energies on the rescaled $l/B$ axis is $1/B$.  Thus $\varphi_{B/2+1} -
\varphi_{B/2} \approx f'(0)$.  But when the actual filling is $\nbar = 
\frac{1}{2}$, we know by particle-hole symmetry that $\varphi_{B/2+1} =
-\varphi_{B/2}$, so $\varphi_{B/2} \approx - f'(0)/2$ and $\varphi_{B/2+1} 
\approx f'(0)/2$.  It follows from \eqref{eqn:dmweight}, that
\begin{equation}\label{eqn:ratio}
\frac{w_{F+1}}{w_F} = \frac{w_{F-1}}{w_F} \approx \exp(-f'(0)).
\end{equation}
For $\nbar \neq \frac{1}{2}$, \eqref{eqn:complete} would tell us that
these ratios are approximately $\exp(-f'(\nbar, 0))$.

This is quite different from what would happen when the `block' contains half
of the entire system, as considered in the standard DMRG algorithm, or in
Ref.~\onlinecite{chung01}.  If the fraction ($B/N$) in the block approached
one, the state $\ket{F}$ must become the Fermi sea ground state of the overall
system, and consequently contains all the weight.  If the block is merely a
finite fraction of the system, we still expect a much larger ratio than
\eqref{eqn:ratio}.  It would be interesting to check the behaviour of the
ratios in \eqref{eqn:ratio} for the case $B, N \to \infty$ with $B/N = 1/2$,
but we have not investigated this.

\subsection{The Gapped Chain of Noninteracting Spinless Fermions}

For the purpose of understanding the pseudo-energy spectrum of non-interacting
systems better, we also considered the dimerized tight-binding Hamiltonian
\begin{equation}\label{eqn:dimerized}
H = -t\sum_{j=1}^N\left[1 + (-1)^j\delta\right]\left(\cxd{j}\cx{j+1} +
\cxd{j+1}\cx{j}\right),
\end{equation}
where the hopping integral $t$ is modulated by the $(-1)^j\delta$ term to
produce an energy gap.  Henceforth we choose the scale of 
energy to be such that $t = 1$.  This system was solved analytically by 
Gebhard \emph{et al},\cite{gebhard97} wherein the Hamiltonian can be written as
\begin{equation}
H = \sum_{|k| < \frac{\pi}{2a}} E(k)\left(a_{k,+}^{\dagger}a_{k,+} - 
a_{k,-}^{\dagger}a_{k,-}\right),
\end{equation}
with $E(k) = \sqrt{\epsilon^2(k) + \Delta^2(k)}$, where $\epsilon(k) = 
-2\cos k$ and $\Delta(k) = 2\delta\sin k$.  In terms of $\Delta(k)$ and 
$\epsilon(k)$, we can define an angle $\phi_k$ such that $\tan 2\phi_k = 
\Delta(k)/\epsilon(k)$, and whose sine and cosine we denote as $\alpha_k = 
\cos\phi_k$, $\beta_k = \sin\phi_k$.  In terms of these, the operators 
$a_{k,+}$ and $a_{k,-}$ for the upper and lower bands respectively are given by
\begin{equation}
\begin{aligned}
a_{k,-} &= \alpha_k\ck{k} + \im\beta_k\ck{k + \pi}, \\
a_{k,+} &= -\beta_k\ck{k} + \im\alpha_k\ck{k + \pi}.
\end{aligned}
\end{equation}

\begin{figure}[htbp]
\centering
\includegraphics[width=0.95\linewidth,clip=true]{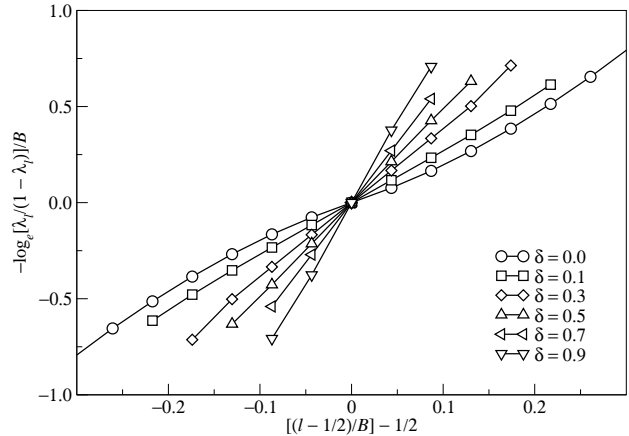}
\caption{Plot of $-B^{-1}\log_e[\lambda_l/(1 - \lambda_l)]$ as a function of
the scaling variable $x = (l-\frac{1}{2})/B - \frac{1}{2}$, for a block of 
size $B = 23$, with different hoping modulations 
$\delta = 0.1, 0.3, 0.5, 0.7, 0.9$.  Pseudo-energies for $|x| > 0.2$ 
for $\delta > 0$ are not shown because these are severely affected by 
numerical errors incurred in the numerical integration and diagonalization 
routines.  The various sets of straight line segments are intended to guide 
the eye in visualizing the data.}
\label{fig:dimerphifixedsize}
\end{figure}

\begin{figure}[htbp]
\centering
\includegraphics[width=0.95\linewidth,clip=true]{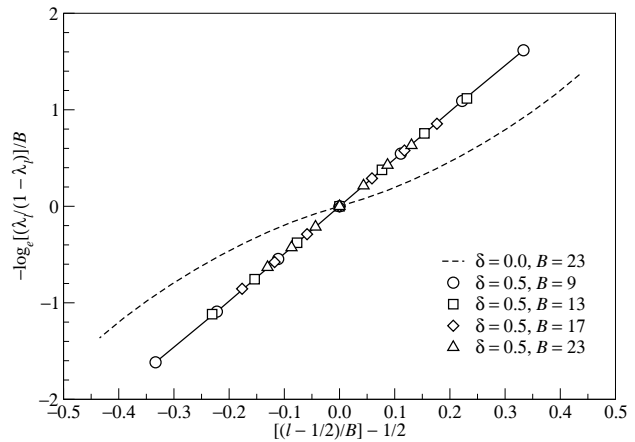}
\caption{Plot of $-B^{-1}\log_e[\lambda_l/(1 - \lambda_l)]$ as a function of
the scaling variable $x = (l-\frac{1}{2})/B - \frac{1}{2}$, for hoping 
modulation $\delta = 0.5$ and different block sizes 
$B = 9, 13, 17, 23$.  The pseudo-energies for $|x| > 0.3$ are not
shown as these are severely affected by numerical errors in the numerical
integration and diagonalization routines.  Also shown, is an approximate 
dashed curve for the scaling function $f(x)$ for $\nbar = \frac{1}{2}$ and
$\delta = 0$, obtained from the data for $B = 23$.  The straight line segments
are intended to guide the eye in visualizing the data.}
\label{fig:dimerphifixeddelta}
\end{figure}

At half-filling, the lower band is completely filled while the upper band is
completely empty, and the ground state can written simply as
\begin{equation}
\ket{\Psi} = \prod_{|k| < \frac{\pi}{2}} a_{k,-}^{\dagger}\ket{0}.
\end{equation}
For this ground state, the two-point function is given by
\begin{multline}
G_{ij} = \frac{1}{2\pi}\int_{-\frac{\pi}{2}}^{\frac{\pi}{2}}dk\,
e^{\im k(i - j)}
\frac{\cos k - \im(-1)^i\delta\sin k}{\sqrt{\cos^2k + \delta^2\sin^2k}},
\end{multline}
using which we can construct the block Green function matrix $G$, and hence
using \eqref{eqn:pseudoenergy} the pseudoenergies which correspond to the
density matrix eigenvalues.  For
a fixed block size of $B = 23$, the pseudo-energy spectra for different hoping 
modulation $\delta$ is shown in FIG.~\ref{fig:dimerphifixedsize}, compared to
that of the gapless case.  Scaling behaviour of the single-particle 
pseudo-energies was found for all $\delta$, each governed by a different
scaling function $f(\nbar, \delta, x)$.  The scaling collapse plot for
$\delta = 0.5$ is shown in FIG.~\ref{fig:dimerphifixeddelta}, compared to
the scaling function $f(\nbar, \delta=0, x)$ for the gapless case.  

Repeating our analysis for the three density matrix eigenstates with the
largest weights, we find again that the ratios of density matrix weights
$w_{F+1}/w_F$ and $w_{F-1}/w_F$ to be independent of block size $B$ when 
the overall system is at half-filling.  However, these ratios 
depend strongly on the hopping modulation $\delta$.  As we can see from
FIG.~\ref{fig:dimerphifixedsize}, the slope of the scaling
curve at $x = 0$ is steeper for larger $\delta$.  This indicates
that --- everything else being equal for finite $B$ --- a smaller fraction of
density matrix states is needed to capture the same total weight if the
system is gapped.

We have not investigated the case $B/N \to 1/2$, as in the standard DMRG
algorithm, but we naturally expect the ratio to increase in a gapped system.
Thus $\ket{F}$ would be a better approximation to the ground state in a
gapped system than in a gapless system, which is known as an empirical fact
in the DMRG context.  Our approach, if extended to the case $B/N > 0$, would
give an analytic justification for this common observation.

\subsection{Largest Density Matrix Weight}

\begin{figure}[htbp]
\centering
\includegraphics[width=0.95\linewidth,clip=true]{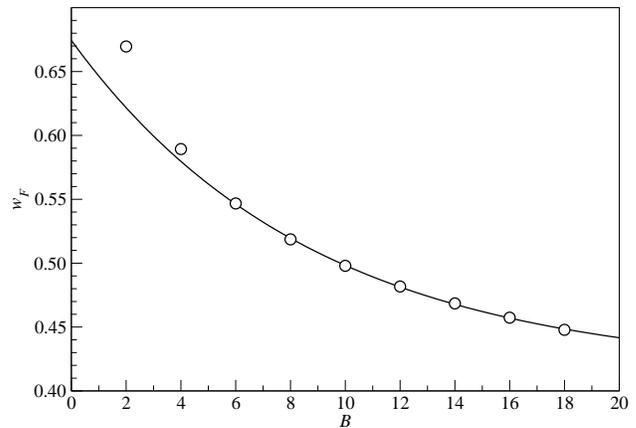}
\caption{Plot of the largest density matrix weight $w_F$ as a function of
the block size $B$ for $B$ even.  The solid curve shown is a fit of the
form $w_F = w_{F,\infty} + \Delta w_F\,\exp(-B/B_0)$.  The best fit to this
small data set is obtained by neglecting the data points for $B = 2$ and
$B = 4$, for which we get $w_{F,\infty} = 0.41$, $\Delta w_F = 0.26$ and $B_0
= 0.11$.}
\label{fig:maxwt}
\end{figure}

For even $B$ blocks on a gapless chain of noninteracting spinless fermions
described the Hamiltonian \eqref{eqn:hopping}, the largest density matrix 
weight $w_F$ can be numerically
computed reliably till $B \approx 20$, and its dependence on $B$ is shown in
FIG.~\ref{fig:maxwt}.  Also shown in FIG.~\ref{fig:maxwt} is a fit of the
numerical data to
\begin{equation}
w_F(B) = w_{F,\infty} + \Delta w_F\,\exp(-B/B_0),
\end{equation}
where $w_{F,\infty}$, $\Delta w_F$ and $B_0$ are curve-fitting parameters.
Here the exponentially decaying term is merely chosen to produce a good curve 
fit --- we believe the $B$-dependence may be more complex --- but what is
interesting is the fact that $w_F$ tends to a constant, $w_{F,\infty}$, in
the limit of $B \to \infty$.  We find that we can understand this in terms of
the scaling formulas developed so far.

From Section~\ref{sect:consistenttruncation} we saw that the largest
many-particle density matrix weight $w_F$ corresponds to the situation for
which all PELs below the pseudo-Fermi level
$\varphi_F = 0$ are occupied, and all those above are empty.  This means that
\begin{equation}\label{eqn:maxweight}
w_F = \mathscr{Q}^{-1}\prod_{l < l_F} e^{-\varphi_l}.
\end{equation}
Using the fact that $\mathscr{Q}$ can be written explicitly as
\begin{equation}
\mathscr{Q} = \prod_l \left(1 + e^{-\varphi_l}\right),
\end{equation}
we then find that
\begin{equation}
w_F = \prod_{l < l_F}\frac{e^{-\varphi_l}}{1 + e^{-\varphi_l}}
\prod_{l > l_F}\frac{1}{1 + e^{-\varphi_l}} = 
\prod_l \frac{1}{1 + e^{-|\varphi_l|}}.
\end{equation}

To evaluate $w_F$, we evaluate first its logarithm, which is
\begin{equation}\label{eqn:logmaxwt}
-\log_e w_F = \sum_l \log_e\left(1 + e^{-|\varphi_l|}\right).
\end{equation}
Here we make two approximations.  Firstly, because of \eqref{eqn:complete}, 
we know that $\varphi_l \propto B$, and so except for a handful of 
single-particle pseudo-energies $\varphi_l$ very near $\varphi_F$, all the
exponentials are very small numbers.  Using the approximation $\log_e(1 + x)
\approx x$ for $x \ll 1$, we write \eqref{eqn:logmaxwt} as
\begin{equation}\label{eqn:logmaxwtapprox}
-\log_e w_F \approx \sum_l e^{-|\varphi_l|}.
\end{equation}
Secondly, we note that because of \eqref{eqn:complete}, single-particle
pseudo-energies far away from $\varphi_F = 0$ will contribute negligibly to the 
above sum.  For $B$ sufficiently large, those single-particle pseudo-energies 
making significant contribution in \eqref{eqn:logmaxwt} will lie within a 
small interval about $l_F$ where a linear approximation of the form
\begin{equation}\label{eqn:linapprox}
\varphi_l \approx B f'(0) \frac{l-l_F}{B} = f'(0)(l - l_F)
\end{equation}
adequately describes the pseudo-dispersion relation.  Substituting
\eqref{eqn:linapprox} into \eqref{eqn:logmaxwtapprox}, we find then that
\begin{equation}
\begin{split}
-\log_e w_F &\approx \sum_{l=1}^B e^{-f'(0)|l - l_F|} \\
&\approx 2\sum_{l > l_F}^{\infty} e^{-f'(0)(l - l_F)}.
\end{split}
\end{equation}
This is a geometric series which we can readily sum to give
\begin{equation}
-\log_e w_F = \frac{2}{1 - \exp(-f'(0))},
\end{equation}
i.e.~the largest density matrix weight $w_F$ is found to approach a constant
value of
\begin{equation}\label{eqn:wfvalue}
w_F = \exp\left[-\frac{2}{1 - \exp(-f'(0))}\right]
\end{equation}
as $B \to \infty$.  From FIG.~\ref{fig:scalingplot} and 
FIG.~\ref{fig:completescaling}, we see that $f'(0) \approx 5$, and so this
asymptotic value of $w_F$ is aproximately 0.13.  This is smaller than
the $w_{F,\infty} = 0.41$ found numerically.

\subsection{Discarded Weight}
\label{sect:residue}

\begin{figure}[htbp]
\centering
\includegraphics[width=0.95\linewidth,clip=true]{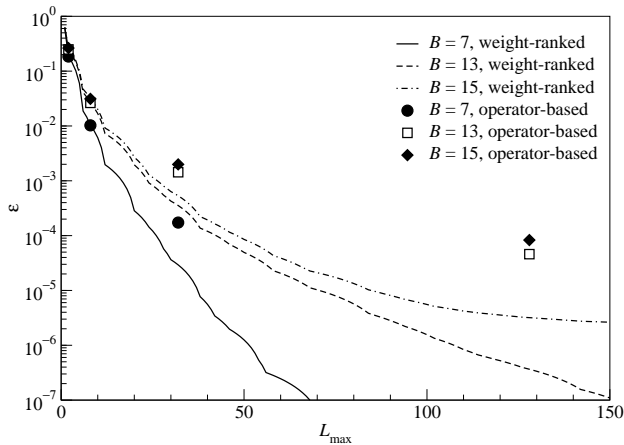}
\caption{Discarded weight $\epsilon$ as a function of the number of states
kept: weight-ranked (as done in the DMRG), and operator-based.}
\label{fig:neps}
\end{figure}

Now that we understand more about the scaling behaviour of the largest density 
matrix weight $w_F$, let us analyze the discarded weight incurred by the 
operator-based DM truncation scheme.  We compute numerically the discarded 
weight incurred by the operator-based DM truncation scheme and that incurred 
by the weight-ranked DM truncation scheme, and show them in FIG.~\ref{fig:neps} 
as a function of the number of many-body states kept as a comparison.  As we 
can see, the discarded weight incurred by the operator-based DM truncation 
scheme is larger compared to the weight-ranked DM truncation scheme, for the 
same number of many-particle eigenstates kept.  This is expected, since the 
weight-ranked DM truncation scheme is by definition the most efficient scheme 
in exhausting the sum rule given in \eqref{eqn:sumrule}.  In spite of this 
seemingly poorer `convergence' property, we believe that the operator-based 
truncation scheme has advantages that cannot be reproduced by the weight-ranked
truncation scheme, to be argued in detail in Section~\ref{sect:cdmts}.

Writing the total density matrix weight explicitly as
\begin{equation}
W = \mathscr{Q}^{-1}\prod_{\text{kept}}(1 + e^{-\varphi_l})
\prod_{\text{below}}(1 + e^{-\varphi_l})
\prod_{\text{above}}(1 + e^{-\varphi_l}),
\end{equation}
where the subscript `kept', `below' and `above' refer to PELs
retained, approximated as always occupied and approximated
as always empty in the operator-based truncation scheme respectively.  The
truncated weight $W_t$ calculated from the operator-based truncation scheme is
\begin{equation}
W_t = \mathscr{Q}^{-1}\prod_{\text{kept}}(1 + e^{-\varphi_l})
\prod_{\text{below}} e^{-\varphi_l}
\prod_{\text{above}} 1.
\end{equation}
Since $W = 1$, the ratio $W_t/W = W_t$ can be written as
\begin{equation}\label{eqn:simpleWt}
\begin{split}
W_t &= \prod_{\text{below}}\frac{e^{-\varphi_l}}{1 + e^{-\varphi_l}}
\prod_{\text{above}}\frac{1}{1 + e^{-\varphi_l}} \\
&= \prod_{\text{below}}\frac{1}{1 + e^{\varphi_l}}
\prod_{\text{above}}\frac{1}{1 + e^{-\varphi_l}}.
\end{split}
\end{equation}

This expression has a simple interpretation in terms of the pseudo-occupation
numbers $\{\lambda_l\}$.  Using \eqref{eqn:pseudooccupationnumber}, we find
that we can write $W_t$ as
\begin{equation}
W_t = \prod_{\text{below}}\lambda_l\prod_{\text{above}}(1 - \lambda_l),
\end{equation}
i.e.~the truncated weight $W_t$ is given by the product of pseudo-occupation
numbers $\lambda_l$ of those PELs we insist 
are always occupied, together with the product of the single-hole 
pseudo-occupation numbers $(1 - \lambda_l)$ of those PELs we insist are 
always empty.  From 
FIG.~\ref{fig:lambdadiff} we see that $\lambda_l$ changes fairly rapidly
from $\lambda_l \lesssim 1$ to $\lambda_l \gtrsim 0$, over a small range of
PELs.  Therefore, it appears that there is
a fairly large range of $l$'s for which $\lambda_l$ is very close to one or
very close to zero.  However, this does not mean that we should perform an
operator-based truncation scheme keeping only the small number of 
PELs whose $\lambda_l$'s are significantly
different from one or zero.  This is because $W_t$ is bounded from above by
\begin{equation}
W_t \leq \prod_{\text{below}}\lambda_{\max}
\prod_{\text{above}}(1 - \lambda_{\min}) \leq (\lambda^*)^{(1 - \gamma)B},
\end{equation}
where $\gamma$ is the fraction of PELs retained
in the operator-based truncation scheme, and
\begin{equation}
\lambda^* = \max(\lambda_{\max}, 1 - \lambda_{\min}).
\end{equation}
Because the exponent is $O(B)$, this number can still be very small.

This brings us to the question we posed in the beginning of this section: how
much of the Hilbert space do we truncate?  If $W_t$ is the only criterion then
we see that a compromise is necessary.  For a small block, the number of
PELs with $\lambda_l$ significantly different
from one or zero is a sizeable fraction of the total number of PELs,
but this number is manageable.  For a large block, the
number of PELs with $\lambda_l$ significantly 
different from one or zero is a tiny fraction of the total number of PELs,
but we still need $\gamma$ to be
reasonably large for $W_t$ to be appreciable in magnitude.  This of course
means that an unmanageably large number of PELs
has to be retained.

To make the above discussions more water-tight, let us make use of the
scaling relations obtained thus far to find a formula relating the truncated
weight $W_t$ to both the block size $B$ and the fraction $\gamma$ of
PELs retained.  Taking the logarithm of
\eqref{eqn:simpleWt} we find, using the fact that $\exp(-|\varphi_l|)
\ll 1$ for $l$ far below $l_F$, and $\exp(-\varphi_l) \ll 1$ for $l$ far
above $l_F$, that
\begin{equation}\label{eqn:almostWt}
\begin{split}
-\log_e W_t &= \sum_{\text{below}} \log_e(1 + e^{\varphi_l}) +
\sum_{\text{above}} \log_e(1 + e^{-\varphi_{l}}) \\
&\approx \sum_{\text{below}} e^{\varphi_{l}} + 
\sum_{\text{above}} e^{-\varphi_{l}} \\
&= \sum_{l < l_F} e^{\varphi_l} - 
\sum_{l = l_F - \gamma B/2}^{l_F} e^{\varphi_l} + {} \\
&\quad\ \sum_{l > l_F} e^{-\varphi_l} -
\sum_{l_F}^{l_F + \gamma B/2} e^{-\varphi_l} \\
&= \Phi^* - \sum_{l = l_F - \gamma B/2}^{l_F} e^{\varphi_l} - {}\\
&\quad\ \sum_{l_F}^{l_F + \gamma B/2} e^{-\varphi_l},
\end{split}
\end{equation}
where $\Phi^*$ is a constant.

If $B$ is large and $\gamma$ small, then the linear approximation
\eqref{eqn:linapprox} for $\varphi_l$ is valid, in which case the two
sums in \eqref{eqn:almostWt} are equal, and given by
\begin{equation}
\begin{split}
\sum_{l_F}^{l_F + \gamma B/2} e^{-\varphi_l} &\approx
\sum_{l_F}^{l_F + \gamma B/2} e^{-f'(0)(l - l_F)} \\
&= \frac{1 - \exp(-\gamma B f'(0)/2)}{1 - \exp(-f'(0))}.
\end{split}
\end{equation}
With this, we can write $W_t$ as
\begin{equation}
W_t \approx W^*\exp\left[\frac{2}{1 - e^{-f'(0)}}
\left(1 - e^{-\gamma B f'(0)/2}\right)\right],
\end{equation}
where $W^* = \exp(-\Phi^*)$.  We can find $W^*$ by taking the limit $\gamma \to
0$, in which case we retain no degree of freedom in the PELs.  
Within the operator-based truncation scheme, this means
that we insist all PELs below $\varphi_F$ to
be always occupied and all those above $\varphi_F$ to be always empty, 
i.e.~only the density matrix eigenstate with the largest weight is retained,
and we should have
\begin{equation}
W_t = w_F = W^*,
\end{equation}
and so
\begin{equation}
W_t \approx w_F\exp\left[\frac{2}{1 - e^{-f'(0)}}
\left(1 - e^{-\gamma B f'(0)/2}\right)\right].
\end{equation}
This can be simplified further, using \eqref{eqn:wfvalue} to get
\begin{equation}
W_t \approx \exp\left[-\frac{2}{1 - e^{-f'(0)}}e^{-\gamma B f'(0)/2}\right].
\end{equation}
In the limit of $\gamma \to 1$, we see from the above expression that
$W_t$ does not tend to one, but we understand that this is because the linear
approximation \eqref{eqn:linapprox} is only valid for a small range of
PELs about $\varphi_F$, i.e.~only for small
$\gamma$.  In this regime, we may further approximate $W_t$ as
\begin{equation}\label{eqn:Wtlmax}
\begin{split}
W_t &\approx \exp\left[-\frac{2}{1 - e^{-f'(0)}}\left(
1 - \frac{f'(0)\gamma B}{2}\right)\right] \\
&\approx w_F \exp\left(\frac{f'(0)}{1 - e^{-f'(0)}}\,l_{\max}\right),
\end{split}
\end{equation}
where
\begin{equation}
l_{\max} = \gamma B
\end{equation}
is the number of PELs retained.  As we can
see, for small $\gamma$, the truncated weight $W_t$ increases exponentially
with $l_{\max}$.  Also, whenever \eqref{eqn:Wtlmax} is valid, we get 
approximately the same truncated weight $W_t$ whether we use $B = 100$ and
$\gamma = 0.2$ or $B = 200$ and $\gamma = 0.1$.  We will see in 
Section~\ref{sect:cdmts} that whenever the retained $\gamma B$ PELs lies 
within the regime where the pseudo-dispersion relation is linear, the
truncation errors are essentially determined by $l_{\max} = \gamma B$.

\section{Single-Particle Density Matrix Eigenfunctions}
\label{sect:wavefunctionscaling}

\subsection{\emph{A Priori} Expectations}

As noted already in Section~\ref{sect:recipe}, in the many-body
eigenstates with largest weights, all the very negative PELs
will be occupied and all the very positive PELs will be empty.  The only 
PELs with significant varying occupancy are those near the pseudo-Fermi
level.

By construction, the many-body density matrix eigenstates with large weights
constitute the likely configurations of the block.  The difference between 
the large-weight eigenstates of the $P$-particle and $(P+1)$-particle
sectors of the density matrix is in the application of a creation operator
$f_l^{\dagger}$ such that the pseudo-energy $|\varphi_l|$ is close to
$\varphi_F = 0$.  In real space, it is likeliest that we can add a particle
near the ends of the $B$-site block, for one can imagine that, in the first
configuration, this particle was just past the end in an adjacent block, and
we merely hopped it a short distance across the boundary to create the state
of $(P+1)$ particles on the block in question.  It follows that the 
single-particle eigenfunctions with single-particle pseudo-energies near the
pseudo-Fermi level have their greatest amplitude near the block's boundaries.
In other words, it is the sites near the end that are most correlated with
the environmental information that we discarded.

\subsection{General Features}

\begin{figure}[htbp]
\centering
\includegraphics[width=\linewidth,clip=true]{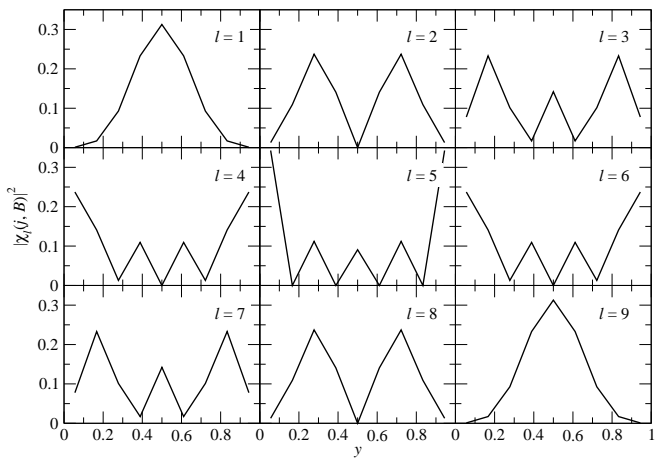}
\caption{Probability density of the normalized one-particle eigenfunctions 
$\chi_l(j)$ (plotted against the scaling variable $y = (j - \frac{1}{2})/B$) 
of $\rho_B$ on a block of $B = 9$ sites at half-filling, showing the 
particle-hole symmetry of the overall system.  The subplots are arranged in 
order of increasing pseudo-energy.}
\label{fig:chik}
\end{figure}

As noted earlier, the eigenstates of $\rho_B$ are all built up from the
eigenstates $\ket{\chi_l}$ of $G$, which are simultaneously the one-particle
eigenstates of $\rho_B$.  As such, the effects of basis truncation,
particularly in obtaining a truncated expansion of the target state 
$\ket{\Psi}$, must be understood in terms of the features of these one-particle
eigenstates.  The real-space features of $\ket{\chi_l}$ can most easily be
understood in terms of the corresponding eigenfunctions $\chi_l(j)$, 
where $j = 1, \dots, B$ are sites on the block.  At half-filling, the 
probability densities $|\chi_l(j)|^2$ exhibit particle-hole symmetry, 
as is shown in FIG.~\ref{fig:chik} for the example of $B = 9$.
In general, by node counting, we see that the sequence of $B$ single-particle
eigenfunctions are in one-to-one correspondence with the sequence of $B$ plane
wave states on the block, where the ordinal number $l$ of the single-particle
eigenfunctions is related to the wavevector $k$ of the plane wave states on
the block by
\begin{equation}
k = \frac{2\pi(l-1)}{Ba}, \qquad l = 1, \dots, B.
\end{equation}

\subsection{Scaling Behaviour}

\begin{figure}[htbp]
\centering
\includegraphics[width=.9\linewidth,clip=true]{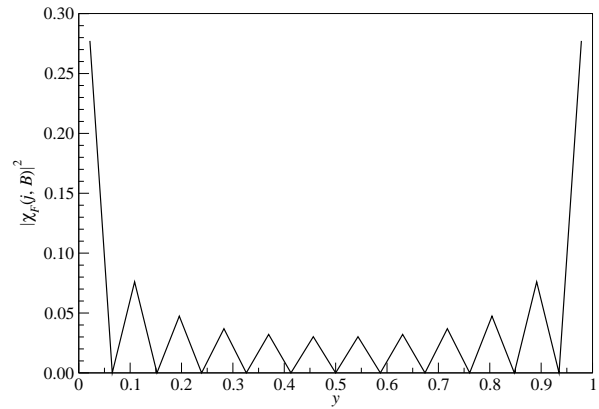}
\caption{Probability density function $|\chi_F(j, B)|^2$ of the pseudo-Fermi 
eigenfunction for $B = 23$, plotted against the scaled variable $(j - 
\frac{1}{2})/B$.}
\label{fig:chif}
\end{figure}

For odd $B$, the pseudo-energy $\varphi_{(B+1)/2}$ sits at the pseudo-Fermi
level, and we may call the corresponding eigenfunction $\chi_F(j) \equiv
\chi_{(B+1)/2}(j)$ the \emph{pseudo-Fermi eigenfunction}.  The probability 
density associated with $\chi_F(j)$ has nodes at every even $j$, as shown in
in FIG.~\ref{fig:chif} for the case of $B = 23$.
The most prominent feature of the pseudo-Fermi eigenfunction, i.e.~the 
amplitude being strongest near the boundaries of the block, was first observed 
by White.\cite{white93}  This appears to be a generic feature that occurs in
both integrable and nonintegrable 1-dimensional systems.  Using the example of
a chain of coupled harmonic oscillators, Gaite explained this ``concentration 
of resolution of quantum states near the boundaries'' as a simple consequence 
of angular quantization of the density matrix.\cite{gaite01}

\begin{figure}[htbp]
\centering
\includegraphics[width=.9\linewidth,clip=true]{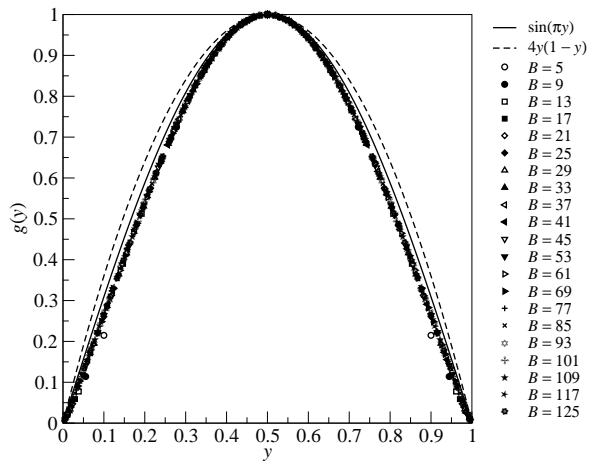}
\caption{Plot of the rescaled envelope function $g(y)$ for various block sizes 
$B = 4p + 1$, $p = 1, 2, \dots$, compared against $\sin\pi y$ and $4y(1 - y)$, 
where $y = (j-\frac{1}{2})/B$ is the rescaled coordinate on the block.}
\label{fig:envelope}
\end{figure}

To analyze $|\chi_F(j,B)|^2$ (where we write the $B$ dependence of $\chi_F(j)$ 
more explicitly) more carefully, we first rescale the eigenvectors obtained
from \textsf{Octave}\cite{octave} such that
\begin{equation}
|\chi_F((B+1)/2, B)|^2 = 1
\end{equation}
for $B = 4p + 1$, $p = 1, 2, \dots$.  For $B = 4p + 3$, $|\chi_F((B+1)/2, B)|^2
= 0$ and the rescaling cannot be carried out as unambiguously as for the
$B = 4p + 1$ series.  This rescaling is harmless, since eigenvectors are
only defined up to an arbitrary normalization.  After this trivial rescaling, 
we find that the pseudo-Fermi probability density can be put in a scaling form
\begin{equation}\label{eqn:pscale}
|\chi_F(j,B)|^2 \cong N(B) g(y)
\frac{\frac{1}{2}[1 - (-1)^j]}{\sin^2\pi y},
\end{equation}
where $y \equiv (j - \frac{1}{2})/B$ and $g(y)$ is the scaling function 
shown in FIG.~\ref{fig:envelope}.

In \eqref{eqn:pscale}, $N(B)$ is a $B$-dependent normalization factor, 
chosen to ensure that the pseudo-Fermi wavefunction given is properly 
normalized in the limit of $B \to \infty$, i.e.
\begin{equation}
\lim_{B\to\infty} \sum_{j=1}^B |\chi_F(j, B)|^2 = 1.
\end{equation}
Although we cannot compute $N(B)$ analytically, we venture a guess to 
its $B$-dependence by noting that the functions $g(y)$ and $\sin\pi y$ 
are not very different from the function $4y(1 - y)$, and so we expect
\begin{equation}
\begin{split}
&\quad\ \sum_{j=1}^B g(y) \frac{\frac{1}{2}[1 - (-1)^j]}{\sin^2\pi y} \\
&\sim \sum_{j=1}^B 4 y (1 - y) 
\frac{\frac{1}{2}[1 - (-1)^j]}{[4 y (1 - y)]^2}\\
&= \frac{1}{4} \sum_{\text{$j$ odd}}^B \left[\frac{1}{y} + 
\frac{1}{1 - y}\right],
\end{split}
\end{equation}
which we can easily work out to have the form
\begin{equation}
\sum_{j=1}^B g(y) \frac{\frac{1}{2}[1 - (-1)^j]}{\sin^2\pi y} \sim
B\left(\log_e B + C\right),
\end{equation}
where $C$ is a constant.  Numerically, the best fit for $N(B)$ in 
the range of block sizes $B = 33$ to $B = 125$ is obtained with
\begin{equation}
N^{-1}(B) = 0.249 B\log_e B + 0.668 B.
\end{equation}

Because of the enhanced amplitude near the edge of the block exhibited in the 
real-space structure of density matrix eigenfunctions with single-particle
pseudo-energies close to the pseudo-Fermi level, and conversely, enhanced
amplitude near the center of the block exhibited in the real-space structure
of density matrix eigenfunctions with single-particle pseudo-energies far
away from the pseudo-Fermi level, we worry that these eigenfunctions might
not be a good basis to use for expanding spatially uniform plane waves,
which are the true single-particle energy eigenstates in our model.  We 
address this concern in Section~\ref{sect:dm2pw}.

\section{Operator-Based Density Matrix Truncation Applied to Dispersion
Relation}
\label{sect:cdmts}

In a gapless system, we conjecture that low-lying excitations \emph{above}
the ground state are built from the same operators as the long-wavelength
fluctuations \emph{within} the ground state.  This supposition is certainly
validated if the system has a continuous symmetry and the long-wavelength
modes are Goldstone modes.  In general it is justified by the relationships
between correlation functions (for the ground-state fluctuations) and
response functions (for low-energy excitations).

Despite its poor convergence properties as far as exhausting the sum rule
\eqref{eqn:sumrule} is concerned, the operator-based truncation would still 
get the salient 
features of the physics right.  We check this by projecting the Hamiltonian
in \eqref{eqn:hopping}
onto the truncated set of fermion operators $f_l$, and calculate the dispersion
relation therefrom.  There are two physical quantities of interest here: (a)
for odd number of sites $B$, the middle band crosses the Fermi level, and we 
can ask how the Fermi velocity, given by the slope of the dispersion relation 
at the Fermi level, scales with $B$ and the fraction $\gamma$ of fermion 
operators kept; or (b) for even $B$, a band gap develops as a
result of truncation at the Fermi level, and we can ask how the size of this
band gap depends on $B$ and $\gamma$.

\subsection{Energy Gap at Fermi Level}

\begin{figure}[htbp]
\centering
\includegraphics[width=.95\linewidth,clip=true]{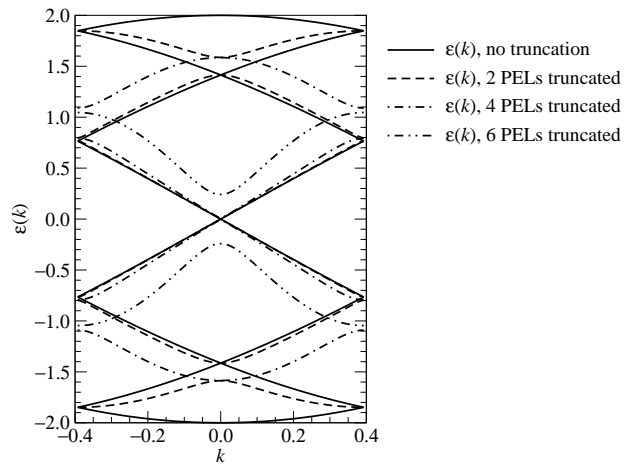}
\caption{Dispersion relation $\epsilon(k)$ for a block of $B = 8$ sites,
where the effects of truncating 2, 4 and 6 PELs are
shown.  For this block size, truncating 2, 4 and 6 PELs 
corresponds to fractions $\gamma = 0.75$, 0.50 and 0.25 of PELs
retained.  For $\gamma = 0.75$, the energy bands (dashed curves) just 
below and above the Fermi level $\epsilon_F = 0$ agree with the true dispersion
relation (solid curve) so well that the difference is not discernible at the 
scales presented in the figure.}
\label{fig:EkB8}
\end{figure}

\begin{figure}[htbp]
\centering
\includegraphics[width=.95\linewidth,clip=true]{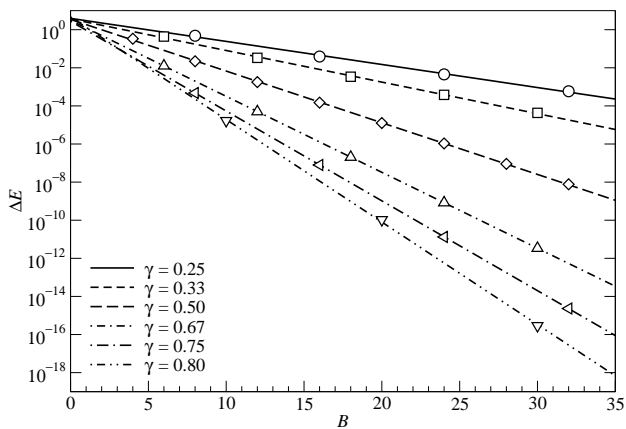}
\caption{Plot of the gap $\Delta E$ as a function of block size $B$ for
various constant fractions $\gamma = 1/4$, $1/3$, $1/2$, $2/3$, $3/4$, $4/5$
of PELs retained.  Also shown are fits to the data points
for various $\gamma$ to $\Delta E(\gamma, B) = \Delta E_0 \exp(-\kappa(\gamma)
B)$.}
\label{fig:gapdelta}
\end{figure}

\begin{figure}[htbp]
\centering
\includegraphics[width=.95\linewidth,clip=true]{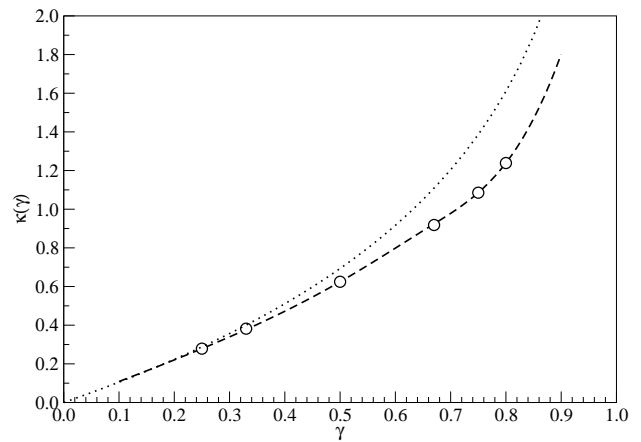}
\caption{Plot of the attenuation coefficient $\kappa(\gamma)$ with which the 
spurious gap $\Delta E$ decays, as a function of the fraction $\gamma$ of PELs
retained.  A cubic spline curve (dashed curve) is superimposed on 
the data points to aid visualization.  Also shown (dotted curve) is $-\log_e(1
- \gamma)$.}
\label{fig:gapdecay}
\end{figure}

In FIG.~\ref{fig:EkB8}, we show the general features of the dispersion
relation $\epsilon(k)$ calculated within the operator-based truncation scheme,
using the example of a block of $B = 8$ sites.  Apart from the energy gap
$\Delta E$ that opens up at the Fermi level $\epsilon_F = 0$, we see that
there is a one-to-one correspondence between the PEL
truncated and the energy band absent from the dispersion relation.  More
precisely, if we order the energy bands and the PELs from
the lowest to the highest as $\{\epsilon_1(k), \dots, \epsilon_B(k)\}$ and
$\{\varphi_1, \dots, \varphi_B\}$, then if we truncate PEL
$\varphi_l$, the energy band $\epsilon_l(k)$ will also be removed from
the numerically calculated dispersion relation.  For fixed
$\gamma$, the gap $\Delta E$ decays exponentially with block size $B$, as
is shown in FIG.~\ref{fig:gapdelta}, i.e.~we have
\begin{equation}\label{eqn:gapdm}
\Delta E = \Delta E_0\exp(-\kappa(\gamma) B),
\end{equation}
where $\kappa(\gamma)$ is an attenuation coefficient whose
$\gamma$-dependence is shown in FIG.~\ref{fig:gapdecay}.  Here we see also 
that $\Delta E(B, \gamma)$ for different $\gamma$ appears to converge onto
$\Delta E_0 \equiv \Delta E(B = 0)$.  Of course, there is no 
physical sense in talking about a block of zero size, but it is nevertheless 
a useful number to keep in mind when studying the scaling behaviour 
of $\Delta E(B, \gamma)$ as $\gamma$ varies.  $\Delta E_0$ is approximately
4, which is the bandwidth of the exact dispersion relation, in all cases.

In particular, in the limit of $\gamma \to 1$, where all PELs 
are retained, the gap is exactly zero for all nonzero block sizes $B$.  In 
this limit, if we start out at a `gap' of $\Delta E_0$ at a `block size' of 
$B = 0$, then to have $\Delta E = 0$ at $B = 1$, we need the attenuation 
coefficient $\kappa$ to be infinite, i.e.~we
expect the limiting behaviour $\lim_{\gamma \to 1} \kappa(\gamma) = \infty$.  
On the other hand, in the limit of $\gamma \to 0$, where we retain none of the
PELs, it is again physically meaningless to talk of a
dispersion relation.  Nevertheless, if we pretend that we are able to 
calculate a `dispersion relation' in this limit, then it is reasonable,
following the trend observed in FIG.~\ref{fig:gapdelta}, that the gap never
closes, i.e.~we expect the limiting behaviour $\lim_{\gamma \to 0} 
\kappa(\gamma) = 0$.  These limiting behaviours appear to be borne out in
the trend observed in FIG.~\ref{fig:gapdecay}.

Another notable feature in FIG.~\ref{fig:gapdecay} is the fact that 
$\kappa(\gamma) \approx \gamma$ for $\gamma \ll 1$, which is the regime
we are most interested to apply the operator-based truncation scheme in.  To
appreciate the relevance of this observation, let us first note from 
FIG.~\ref{fig:EkB8} the general feature that the smaller the gap $\Delta E$,
the better the truncated dispersion relation matches the true dispersion
relation about the Fermi level.  From \eqref{eqn:Wtlmax} we saw that the
truncated weight $W_t$ depends only on the combination $l_{\max} = \gamma B$
in this regime, and as far as $W_t$ is concerned, there is no difference 
whether we choose to keep 10 out of 100 PELs ($\gamma = 0.1$) or 10 out of 
200 PELs ($\gamma = 0.05$).  Here we see a similar exponential dependence
on $l_{\max}$ for the spurious gap $\Delta E$ that arises due to truncation: 
if we write $\kappa \approx \gamma$ in this regime, then $\Delta E \approx 
\Delta E_0\exp(-B\gamma) = \Delta E_0\exp(-l_{\max})$.  Such an exponential
behaviour implies that we have very good control over the numerical accuracy
of the dispersion relations --- in particular near the Fermi level ----
calculated in the operator-based DM truncation scheme.

\subsection{Fermi Velocity}

\begin{figure}[htbp]
\centering
\includegraphics[width=.95\linewidth,clip=true]{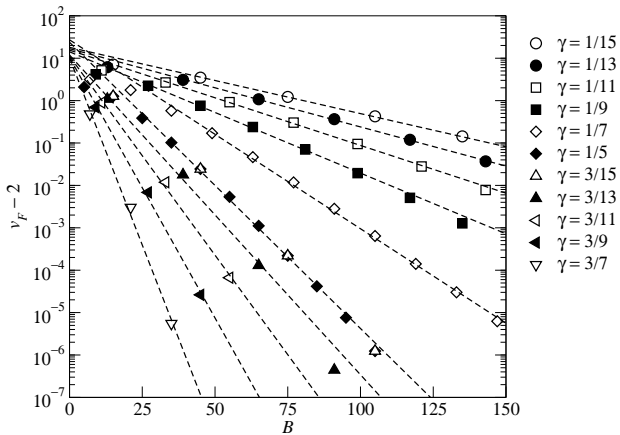}
\caption{Plot of Fermi velocity deviation $(v_F - 2)$ calculated from the 
truncated dispersion relation, as a function of the block size $B$, for various
fractions $\gamma = 1/15, 1/13, 1/11, 1/9, 1/7, 1/5, 3/15, 3/13, 3/11, 3/9,
3/7$ of PELs retained.  Fits to average exponential decays are also shown.}
\label{fig:vfdecay}
\end{figure}

When the block size $B$ is odd, the central energy band crosses the Fermi
level, and the quantity of interest becomes the Fermi velocity $v_F$.  This
can be determined from the truncated dispersion relation by taking the
numerical central derivative of the central energy band at $k = \pm\nbar\pi/B$.
At half-filling, $\epsilon(k = \pm\pi/2B) = 0$ exactly because of particle-hole 
symmetry.  This feature of the dispersion relation was found to be preserved
in the numerical dispersion relations computed within the operator-based DM
truncation scheme.  On the global scale, we find numerically that the shifts 
in the 
central energy band at the Brillouin zone center and Brillouin zone edge are 
such that $v_F > 2$ always.  However, when $B$ is large, the numerical 
diagonalization routine introduces artefacts on the energy scale of 
$10^{-13}$, resulting in the locally evaluated $v_F$ coming out to be very
slightly less than 2.  As such, we analyze the behaviour of $v_F$ as a function
of $B$ and $\gamma$ only for $B < 150$, as shown in FIG.~\ref{fig:vfdecay}.

\begin{figure}[htbp]
\centering
\includegraphics[width=.95\linewidth,clip=true]{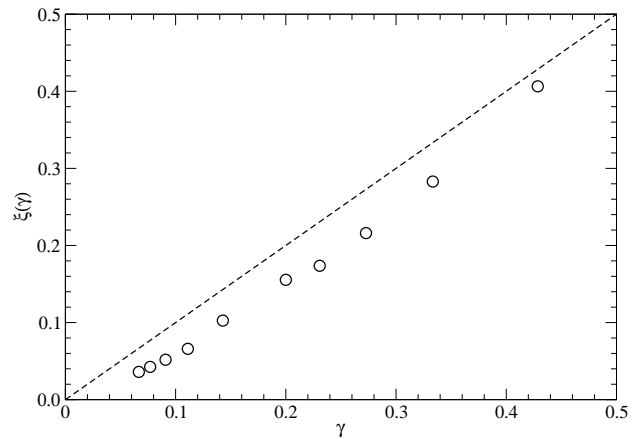}
\caption{Plot of the attenuation coefficient $\xi(\gamma)$ as a function
of the fraction $\gamma$ of PELs retained.
Also shown (dashed line) is the expected behaviour $\xi(\gamma) = \gamma$.}
\label{fig:crossover}
\end{figure}

As can be seen from FIG.~\ref{fig:vfdecay}, the difference $(v_F - 2)$
decays more or less exponentially with $B$ for various $\gamma$, i.e.
\begin{equation}
(v_F - 2) \cong \exp(-\xi(\gamma)B),
\end{equation}
where $\xi(\gamma)$ is the $\gamma$-dependent attenuation coefficient for
the average exponential decay.  The $\gamma$-dependence of $\xi(\gamma)$
is shown in FIG.~\ref{fig:crossover}.

\subsection{Real-Space Structure of Eigenfunctions of Truncated Hamiltonian}
\label{sect:dm2pw}

The eigenfunctions of the untruncated Hamiltonian \eqref{eqn:hopping}
are spatially uniform plane waves, with amplitude $\exp(\im kj)/\sqrt{B}$ on
site $j$ of the block of $B$ sites.  These can be expanded in terms of the
density matrix eigenfunctions $\chi_l(j, B)$.  Naively, we expect that if we 
drop those $\chi_l(j, B)$ associated with pseudo-energies $\varphi(l, B)$ far
from the pseudo-Fermi level $\varphi_F$, as we would in our operator-based
truncation scheme which removes these single-particle pseudo-energy levels
as degrees of freedom, the remaining terms, all having enhanced amplitudes
at the edge of the block, would sum to a function with enhanced amplitude at
the edge of the block.  It would therefore seem like we are attempting to
approximate a spatially-uniform plane wave with a function with the wrong
real-space structure.

\begin{figure}[htbp]
\centering
\includegraphics[width=.95\linewidth,clip=true]{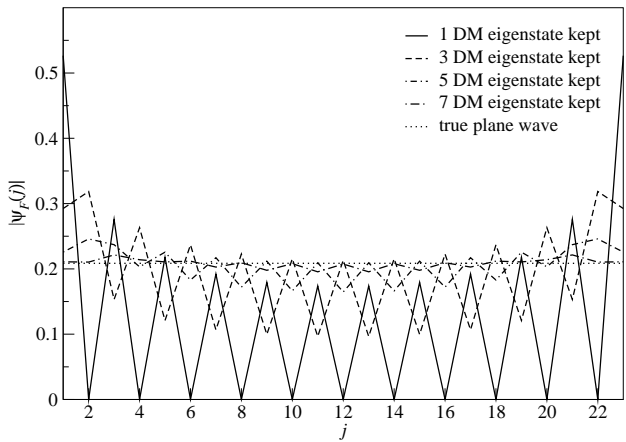}
\caption{Amplitudes of eigenfunctions of the truncated Hamiltonian 
$\tilde{H}(k_F)$ for different numbers of PELs retained, for a block of
$B = 23$ sites at half-filling.}
\label{fig:dm2pw}
\end{figure}

However, the key insight we gain from our study of block density matrices
is that while the system-wide density matrix $\rho_0$ commutes with
$H$ in \eqref{eqn:hopping}, the block density matrix $\rho_B$ obtained
by tracing down $\rho_0$ does not commute with $H(k)$, for all $k$.  
Therefore, after operator-based truncation $H(k) \to \tilde{H}(k)$, we would
need to diagonalize $\tilde{H}(k)$ to find the truncated dispersion relation
$\epsilon_l(k)$.  Thus, the function that would approximate the plane wave
is not the latter's truncated expansion in terms of the eigenfunction of
the one-particle block density matrix, but rather, a particular eigenfunction of
$\tilde{H}(k)$, which is an appropriate linear combination of the
$\chi_l(j, B)$ retained in the operator-based truncation scheme.  We show in
FIG.~\ref{fig:dm2pw} the spatial structure of such a function, for various
numbers of density matrix eigenfunctions retained.  As we can see, for a
block of $B = 23$ sites, keeping 7 density matrix eigenfunctions with
pseudo-energies around $\varphi_F$ would produce a decent approximation to
the plane wave with $k = \pi/2B$.

\subsection{Discussion}

We speculate that the fact that the operator-based density matrix truncation
scheme succeeds so well suggests that appropriate linear combinations of the
density matrix eigenfunctions can closely approximate a plane wave with
wavevector $\pm k_F$.  This is only possible by taking the difference of two
eigenfunctions so as to cancel the enhancements of the envelope function 
seen at the block ends (see Section~\ref{sect:wavefunctionscaling}).  Indeed,
the fact that we get the correct slope $v_F$ of the dispersion near $k_F$
suggests that by taking different weights, a continuously varying effective
wavevector can be approximated.

The fact that the goodness of approximation depends only on the number of
eigenfunctions kept, means that we approximate the wavefunction about as well
in two successive blocks of $B$ sites, as we do in one big block of $2B$
sites.  One could speculate that there might exist some sort of approximate
composition formula, analogous to Clebsh-Gordan formulas for combining
angular momenta, that provides the $2B$-site eigenfunctions in terms of the
direct product of the $B$-site eigenfunctions.

\section{Operator-Based Plane Wave Truncation Scheme}
\label{sect:cpwts}

As we saw in Section~\ref{sect:wavefunctionscaling}, eigenstates of
the density matrix $\rho_B$ are approximately plane waves (with wavevector 
$Q$ determined
by the boundary conditions on the block of $B$ sites) modified by some
envelope function.  Apart from the effects of the envelope functions, the
operator-based truncation scheme described above is likened to truncating
wavevectors $Q$ far away from the Fermi wavevector $k_F$.  It is therefore
natural to investigate how a operator-based truncation scheme based on plane
waves would fare against that based on the density matrix eigenstates.

\subsection{Exact Dispersion at Zone Center}

Compared to the \emph{operator-based DM truncation scheme} developed above,
the most striking feature of the \emph{operator-based plane wave (PW) truncation
scheme} is that it gets the dispersion exactly right at the zone center, as
shown for the case of $B = 8$ in FIG.~\ref{fig:pwdispersiongapless}, and
for the case of $B = 10$ in FIG.~\ref{fig:pwdispersiongapped}.  We understand
this as follows:

\begin{figure}[htbp]
\centering
\includegraphics[width=.95\linewidth,clip=true]{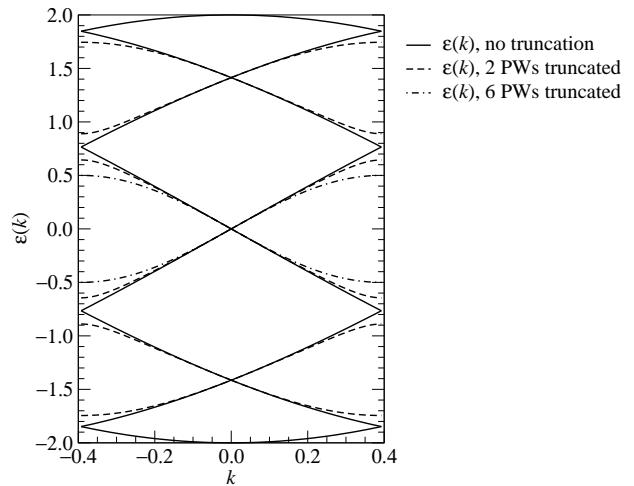}
\caption{Dispersion relation $\epsilon(k)$ for a block of $B = 8$ sites,
where the effect of truncating 2 and 6 plane waves (PWs) are shown.  For
$B = 4n$, the Fermi level is located at the zone center, and $\epsilon(k)$ is
always gapless here regardless of the number of plane waves truncated.}
\label{fig:pwdispersiongapless}
\end{figure}

To evaluate the dispersion relation in a blocked description, we start by
defining the direct Bloch basis states
\begin{equation}\label{eqn:jk}
\ket{j, k} = \frac{1}{\sqrt{N/B}}\sum_J e^{\im kJB}\ket{j, J},
\quad j = 1, \dots, B,
\end{equation}
where $\ket{j, J} = \cxd{j + JB}\ket{0}$ is the single-particle occupation
number basis state at site
$j + JB$ along the chain.  In this basis, the Hamiltonian \eqref{eqn:hopping}
for a chain of $N$ non-interacting spinless fermions take on a block diagonal
form.  Diagonalizing the $B\times B$ diagonal block
\begin{equation}
H(k) = \begin{bmatrix}
0 & -1 & 0 & \cdots & -e^{-\im kJB} \\
-1 & 0 & -1 & \cdots & 0 \\
0 & -1 & 0 & \cdots & 0 \\
\vdots & & & \ddots & \vdots \\
-e^{\im kJB} & 0 & 0 & \cdots & 0 \end{bmatrix}
\end{equation}
for $-\pi/B \leq k < \pi/B$ then gives the dispersion relation within the
reduced zone scheme.

For the operator-based PW truncation scheme, we need to work with the plane wave
states $\ket{Q, J}$ on each block of $B$ sites, where the wavevector $Q$ is 
determined by periodic boundary condition, i.e.~$\exp(\im QB) = 1$.  These 
plane wave states are related to the single-particle occupation number basis
states by
\begin{equation}\label{eqn:QJ}
\ket{Q, J} = \frac{1}{\sqrt{B}}\sum_{j=1}^B e^{\im Qj}\ket{j, J}.
\end{equation}
A Bloch basis state parallel to \eqref{eqn:jk} can be defined as
\begin{equation}\label{eqn:Qk}
\ket{Q, k} = \frac{1}{\sqrt{N/B}}\sum_J e^{\im kJB}\ket{Q, J},
\end{equation}
where $QB/2\pi = 0, \dots, B-1$.
From \eqref{eqn:QJ} and \eqref{eqn:Qk}, it is easy to see that
\begin{equation}
\ket{Q, k} = \frac{1}{\sqrt{B}}\sum_{j=1}^B e^{\im Qj}\ket{j, k}.
\end{equation}

At the zone center, $k = 0$, and the $B\times B$ Hamiltonian matrix in the
$\ket{j, k}$ basis that we need to diagonalize becomes
\begin{equation}
H(0) = \begin{bmatrix}
0 & -1 & 0 & \cdots & -1 \\
-1 & 0 & -1 & \cdots & 0 \\
0 & -1 & 0 & \cdots & 0 \\
\vdots & \vdots & \vdots & \ddots & \vdots \\
-1 & 0 & 0 & \cdots & 0 \end{bmatrix}.
\end{equation}
It is trivial to check that the eigenstates of this Hamiltonian matrix are
precisely the plane waves $\ket{Q, 0}$ on the block.  Therefore, in the
$\ket{Q, k}$ basis, $H(k)$ is diagonal at $k = 0$, and so truncating some
plane waves from the Hilbert space produces no effect on the dispersion here.

To be more precise, in performing truncation, a linear subspace of the Hilbert 
space is chosen, and the Hamiltonian projected onto this subspace.  If 
$\ket{\psi}$ is an eigenstate of the full Hamiltonian, and if $\ket{\psi}$
is retained in the truncated Hilbert space, then it will continue to be an
eigenstate of the truncated Hamiltonian, with the same eigenvalue.

\subsection{Energy Gap at Zone Boundary}

\begin{figure}[htbp]
\centering
\includegraphics[width=.95\linewidth,clip=true]{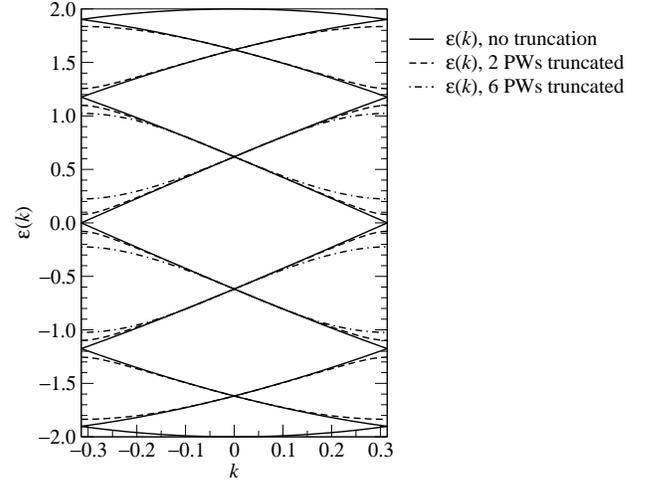}
\caption{Dispersion relation $\epsilon(k)$ for a block of $B = 10$ sites, 
where the effect of truncating 2 and 6 plane waves (PWs) are shown.  For
$B = 4n + 2$, the Fermi level is located at the zone boundary, and the
operator-based plane wave truncation scheme introduces an energy gap $\Delta E$
here.}
\label{fig:pwdispersiongapped}
\end{figure}

For even block sizes with $B = 4n$, the Fermi level is located at the zone
center in the reduced zone scheme, and so there is no energy gap to speak of.
On the other hand, for even block sizes with $B = 4n + 2$, the Fermi level is 
located at the zone boundary.  At the zone boundary, operator-based PW 
truncation introduces an energy gap $\Delta E$ at the Fermi level, as shown in 
FIG.~\ref{fig:pwdispersiongapped} for $B = 10$.

\begin{figure}[htbp]
\centering
\includegraphics[width=.95\linewidth,clip=true]{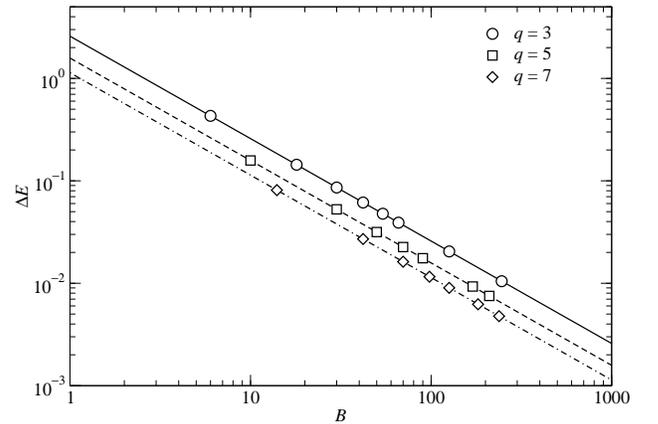}
\caption{Plot of $\Delta E$ as a function of block size $B$ for $q = 3$, 5 and
9, corresponding to the fractions $\gamma = 2/3, 4/5, 6/7$ of block plane 
wave states retained.  Also shown are the fits of the data points to the
formula $\Delta E(B, \gamma) = \Delta E_1(\gamma)/B$.  From the fits, we
have $\Delta E_1 = 2.58263$ for $\gamma = 2/3$, $\Delta E_1 = 1.57991$ for
$\gamma = 4/5$ and $\Delta E_1 = 1.13532$ for $\gamma = 6/7$.}
\label{fig:pwgap}
\end{figure}

As in the case for the operator-based DM truncation scheme, we
investigate the behaviour of the energy gap $\Delta E$ as a function of the 
block size $B$ for a fixed fraction $(1 - \gamma)$ of block states truncated.
However, for the operator-based PW truncation scheme, the number of plane
wave states that can be truncated, if $B$ is even, is $4m + 2$, $m = 0$, 1, 
2, \dots.  Thus the only realizable series of block sizes $B$ on which we 
can perform fixed $(1 - \gamma)$ truncation are of the form $B = q(4m + 2)$,
$q = 2$, 3, \dots.  The fraction $\gamma$ of block plane wave states retained
is related to the series index $q$ by
\begin{equation}
\gamma = 1 - \frac{1}{q}.
\end{equation}
Half of these realizable series have block sizes that
are multiples of 4, for which the Fermi level is at the Brillouin zone center
where the dispersion relation we have shown in the previous subsection to be
gapless.  In this subsection we are interested in those block sizes for which
$q$ is an odd integer, since for these block sizes the Fermi level is at
the Brillouin zone boundary, where a gap develops in the dispersion relation
as a result of truncation.  The behaviour of $\Delta E$ as a function of $B$ 
for three series of $q$ is shown in FIG.~\ref{fig:pwgap}.

As can be seen from FIG.~\ref{fig:pwgap}, the gap depends on block size as an
inverse power law
\begin{equation}
\Delta E(B, \gamma) = \frac{\Delta E_1(\gamma)}{B},
\end{equation}
where $\Delta E_1(\gamma)$ is a $\gamma$-dependent prefactor.  This is in 
stark contrast to the exponential dependence \eqref{eqn:gapdm} found for 
the case of the operator-based density matrix truncation scheme.

\subsection{Fermi Velocity}

\begin{figure}[htbp]
\centering
\includegraphics[width=.95\linewidth,clip=true]{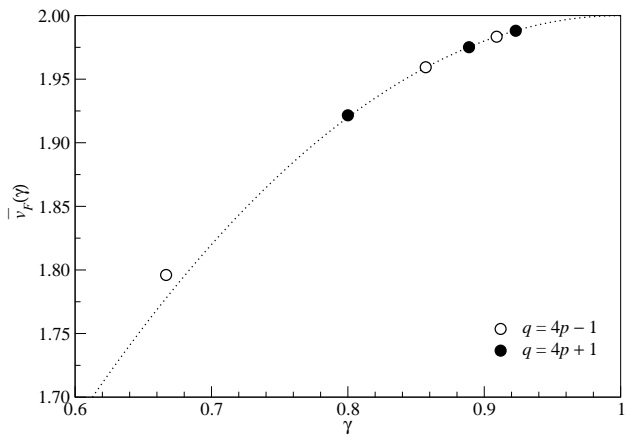}
\caption{Plot of $\bar{v}_F$ as a function of $\gamma$ in the operator-based PW 
truncation scheme for both the $q = 4p-1$ series and $q = 4p+1$ series.  Also
shown as the dashed curve is $2[1-(1-\gamma)^2]$, which appears to fit
the data points well near $\gamma = 1$.}
\label{fig:pwvftilde}
\end{figure}

\begin{figure}[htbp]
\centering
\includegraphics[width=.95\linewidth,clip=true]{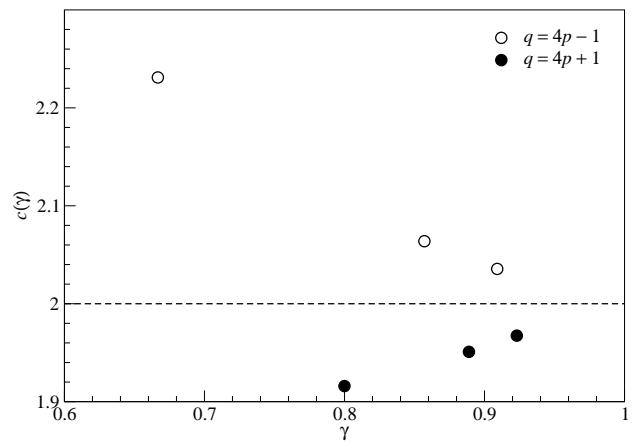}
\caption{Plot of $c_{\pm}$ as a function of $\gamma$ in the operator-based PW
truncation scheme for both the $q = 4p-1$ series ($c_+(\gamma)$) and
$q = 4p+1$ ($c_-(\gamma)$) series.  Both $c_+(\gamma)$ and $c_-(\gamma)$
appears to be converging towards 2.  In fact, from the graph we find that
$|c_{\pm}(\gamma)-2| \approx 0.43(1 - \gamma)$.}
\label{fig:pwc}
\end{figure}

For odd $B$, we again investigate the behaviour of $v_F$ as a function of $B$
for the operator-based plane wave truncation scheme.  The number of block plane
waves that can be truncated is $4m + 3$, $m = 0$, 1, 2, \dots and the series 
of realizable block sizes are $B = q(4m + 3)$, $q = 3$, 5, \dots.  
Unlike in the operator-based DM truncation scheme, there appears to be two 
different systematic behaviours for $v_F(B, \gamma)$, one for $q = 4p - 1$ and 
another for $q = 4p + 1$ ($p = 1$, 2, \dots).  We find that the Fermi velocity
can be fitted very well to the formula
\begin{equation}
v_F = \begin{cases}
\bar{v}_F(\gamma) + c_+(\gamma)/B, & q = 4p - 1; \\
\bar{v}_F(\gamma) - c_-(\gamma)/B, & q = 4p + 1.
\end{cases}
\end{equation}
The plots of $\bar{v}_F(\gamma)$ and $c_{\pm}(\gamma)$ as a function of 
$\gamma$
are shown in FIG.~\ref{fig:pwvftilde} and FIG.~\ref{fig:pwc} respectively.

As can be seen from FIG.~\ref{fig:pwvftilde}, the exact value of the
Fermi velocity is obtained only in the double limit of $B \to \infty$ and
$\gamma \to 1$.  Compared to the operator-based DM truncation scheme, where we
manage to achieve the exact Fermi velocity for any $\gamma$, this is
clearly undesirable.  Furthermore, even very close to $\gamma = 1$, the
computed Fermi velocity approaches the limiting value $\bar{v}_F(\gamma)$ as 
$B^{-1}$.  This is much slower than the exponential convergence of 
$\exp(-\xi(\gamma) B)$ found for the operator-based DM truncation scheme.

\section{Summary and Discussions}
\label{sect:conclusions}

\subsection{Statistical-Mechanical Analogy and Scaling}

To summarize, we have in this paper developed an in-depth understanding of
the structure of the eigenvalues and eigenstates of the density matrix
$\rho_B$ of a block of $B$ sites embedded within a infinite one-dimensional 
chain of noninteracting spinless fermions described by the set of fermion
operators $\{\ck{k}\}$ with dispersion relation $\epsilon(k)$.

From Ref.~\onlinecite{cheong02} we know that the block density matrix $\rho_B$ 
can be written in \eqref{eqn:smform} as the exponential of a pseudo-Hamiltonian 
$\tilde{H}$ given by \eqref{eqn:pseudoH} acting only within the block, which 
describes another system of noninteracting spinless fermions with fermion 
operators $\{f_l\}_{l=1}^B$ and dispersion relation $\varphi_l$.  We use the 
prefix \emph{pseudo-} when talking about operators $\{f_l\}_{l=1}^B$ and 
energies $\varphi_l$ to distinguish them from the operators $\{\ck{k}\}$ and 
energies $\epsilon(k)$ of the system we started out with.

The many-particle eigenstates of $\rho_B$ can then be interpreted simply as
the many-body energy eigenstates \eqref{eqn:dmeigenstate} of the system of 
noninteracting spinless fermions described by $\tilde{H}$, and their
associated density matrix weights \eqref{eqn:dmweight} are then their 
statistical weights in the grand canonical ensemble, \emph{as though} the
block is at a finite temperature.  Within this statistical-mechanical 
picture, we can apply intuitions learnt from the statistical mechanics of
real fermionic systems, and talk about the filling of single-particle
pseudo-energy levels (PELs) as dictated by the Fermi-Dirac distribution, 
the pseudo-Fermi sea with its pseudo-Fermi level $\varphi_F$ and particle-hole 
excitations of the pseudo-Fermi sea.

Using results obtained in 
Ref.~\onlinecite{cheong02}, we make this statistical-mechanical picture
more precise, by identifying the single-particle eigenstates $\ket{\chi_l}$
in \eqref{eqn:chil} of $\rho_B$ as those built up from the eigenvectors
$\{\chi_l(j)\}$, $j = 1, \dots, B$, of the Green function matrix $G$
restricted to the block.  The eigenvalue $\lambda_l$ of $G$, related to
the single-particle pseudo-energies $\varphi_l$ by \eqref{eqn:pseudoenergy}
and \eqref{eqn:pseudooccupationnumber}, are the average pseudo-occupation 
numbers of the $l$\textsuperscript{th} PEL.

The statistical mechanics of real fermionic systems tells us that, at finite 
temperature, the physically important many-body states are those low-energy
particle-hole excitations involving single-particle energy levels within
$k_BT$ of the Fermi level.  Single-particle energy levels far away from
the Fermi level contribute negligibly to the thermodynamic properties of
the fermionic system, and are precisely the degrees of freedom to be
truncated in a renormalization group analysis.

We capitalize on this insight,
and described a recipe for operator-based truncation of the density matrix
eigenstates, where out of the $B$ pseudo-fermion operators $\{f_l\}_{l=1}^B$,
we truncate those $f_l$'s associated with single-particle pseudo-energies 
$\varphi_l$ far away from $\varphi_F$ and retain $l_{\max}$ of them with
$\varphi_l \approx \varphi_F$.  Within this operator-based truncation scheme,
the effective pseudo-Hamiltonian acting on the truncated Hilbert space can 
be made to have the same form as the original pseudo-Hamiltonian, in the
same spirit of renormalization group transformations in statistical
mechanics or quantum field theory.

Having laid out the basic principles behind our operator-based DM
truncation scheme, we proceeded to look more closely into the distribution
of single-particle pseudo-energies $\varphi_l$, and how these scale with the
block size $B$.  There are two related questions that provide the motivation
for doing this: (1) the statistical mechanics of real fermionic systems 
suggests that single-particle energy levels within $\beta^{-1} = k_BT$ of the 
Fermi level are the physically relevant degrees of freedom --- what then is 
the effective temperature $\tilde{\beta}^{-1}$ that we should use as the 
natural cutoff when performing operator-based truncation on $\{f_l\}_{l=1}^B$? 
(2) although we have associated the pseudo-dispersion relation $\varphi_l$ 
with the dispersion relation $\epsilon(k)$ of a real fermionic system of 
noninteracting spinless fermions, the wavevector $k$ enumerating $\epsilon(k)$ 
is an \emph{intensive} quantity whereas the ordinal number $l$ enumerating 
$\varphi_l$ is an \emph{extensive} quantity --- what would the intensive 
analog of $l$ that more closely parallels the wavevector $k$, and how would 
the pseudo-dispersion relation look like in terms of this intensive label?  

The natural answer to the second question would be to write the 
pseudo-dispersion relation $\varphi_l$ as a function of $l/B$, totally 
analogous with how the wavevector $k$ is enumerated as $2\pi m/N$ for a chain 
of $N$ sites satisfying the Born-von Karman boundary condition.  In fact, we 
find strong numerical evidence that suggests that the single-particle 
pseudo-energies, for various block sizes and fillings, satisfy a scaling 
relation of the form given in \eqref{eqn:complete},
where the scaling function $f(\nbar, x)$ is the proper analog of the
dispersion relation $\epsilon(k)$, and the scaling variable $x$ given in
\eqref{eqn:scalingvariable} is the proper analog of the wavevector $k$.  From
\eqref{eqn:lambdascale}, we see that the block size $B$ plays the role of 
inverse temperature.

Our scaling results in Section~\ref{sect:scaling} indicate that the density
matrix eigenstates and eigenvalues behave, as block size $B$ is increased,
very much as energy eigenstates and eigenvalues do when the system size is
increased.  In the latter case, we have a dispersion relation and are more
or less sampling it at different wavevectors.  It is not quite that simple
in the density matrix case, since the scaling function \eqref{eqn:complete}
--- our analog of the energy dispersion relation --- depends on the filling
$\nbar$.  We only note that this analogy still lacks an analytical foundation.
A more penetrating analysis is called for of the relation of $G$ to $\rho_B$
(or equivalently, the effect on its eigenvalues of restricting $G$ to sites
on a block).

Incidentally, we noted that our equation relating $G$ to $\rho_B$ was valid at 
any temperature, but we assumed zero temperature throughout this paper.  We 
expect nonzero temperature $T$ would become a second scaling variable.  Since 
$T > 0$ has similar effects on the Green function $G(r)$ as does the gap 
introduced in \eqref{eqn:dimerized}, we expect the scaling also behaves 
similarly and we did not investigate it.

\subsection{Scaling as a Guide to Truncation in Practice}

In a real application, it appears quite unlikely that Hamiltonians will be
projected directly onto large blocks (meaning blocks of more than 16 sites).
What then is the practical value of extracting scaling forms, if they are
unambiguously seen only in blocks of 100 or more sites?  One
answer is that, even though it is an oversimplified cartoon for the
non-asymptotic situations in which it usually gets applied, a scaling law
is easier to grasp than brute numerical or graphical facts.

The scaling relation \eqref{eqn:complete} is a powerful tool that we can use
to derive deeper understanding concerning the structure of the block density
matrix, as well as various aspects of truncation.  But in itself, the scaling 
relation provides only a partial answer to our first question, which is about 
how much of the Hilbert space of many-body states on the block of $B$ sites 
we can truncate.  To answer this question more completely, we looked at the
three density matrix eigenstates $\ket{F}$, $\ket{F-1}$ and $\ket{F+1}$ with
the largest weights.  Using the scaling relation \eqref{eqn:complete}, we
find that the ratios $w_{F\pm 1}/w_F$ of their weights ($w_F$ being the
largest density matrix weight) approach a constant, in \eqref{eqn:ratio} for
a gapless system of noninteracting spinless fermions, as $B \to \infty$.

The same result was also found for a gapped system of noninteracting spinless 
fermions, for which we find scaling relations governed by gap-dependent
scaling functions.  Furthermore, the scaling relation \eqref{eqn:complete}
allowed us to conclude that as $B \to \infty$, the largest density matrix 
weight $w_F$ approaches a constant given in \eqref{eqn:wfvalue}, and derive
approximately the dependence on the number $l_{\max}$ of PELs retained and
the truncated weight $W_t$ in the operator-based DM truncation scheme.

\subsection{DMRG and Operator-Based Truncation}

It is difficult to compare our results with those obtained in the context
of the DMRG, because that is an incremental method.  Rather than obtain the
density matrix of a large block all at once, each iteration of the DMRG takes
an approximate density matrix for a block of $B$ sites and produces an
approximate density matrix for a block of $(B+1)$ sites.  The fraction of
weight kept, which is taken as the figure of merit, refers to the small
truncation in each iteration.  The cumulative DMRG truncation might be more
appropriate to compare with our results for rather large blocks.

Nevertheless, let us note that operator-based truncation can be applied 
independent of whether we use an incremental or one-shot method.  In 
particular, operator-based truncation could be used in a test run to apply 
DMRG to a noninteracting Fermi chain.  One is given a truncated list of $t$ 
operators $\{f_l\}$ where $l = (B - t + 1)/2, \dots, (B + a - 1)/2$ for the 
original block, and a Hamiltonian projected onto it.  One augments this list 
with the bare creation operators $c_{B+1}^{\dagger}$ on a new site that will 
be added, and defines the new Hamiltonian by adding the hopping to the new 
site.

In light of the derivation of Ref.~\onlinecite{cheong02}, we anticipate 
that the density matrix of the augmented system's ground state must have the
same quadratic form, with new operators $\{f'_l\}$, which could presumably
be obtained merely by diagonalizing the single-particle sector.  One simply
deletes the least important member of this list to obtain a new truncated
list, no longer than the original one.

This difficulty notwithstanding, we still carried out a naive comparison of 
the performance of the operator-based DM truncation scheme against the 
traditional weight-ranked DM truncation scheme used in the DMRG, using
the ability to exhaust the sum rule \eqref{eqn:sumrule} for a given total
numer $L_{\max}$ of density matrix eigenstates retained as a criterion.
The conclusion: while the operator-based DM truncation scheme do not exhaust
the sum rule \eqref{eqn:sumrule} as rapidly as the weight-ranked DM truncation
scheme, $W_t$ is still of $O(1)$, i.e.~the significant parts of the total 
density matrix weight are `captured' by the operator-base DM truncation scheme.

\subsubsection{Truncation and Dispersion Relations}

However, we believe it is more important to check how well a truncation scheme 
do by calculating physical quantities, rather than rely solely on the truncated 
weight $W_t$ as a performance indicator.  To this end, we calculated the 
dispersion relation of elementary excitations within the operator-based DM 
truncation scheme (an easy thing to do), and found that the error incurred 
decays exponentially as $l_{\max}$, the number of PELs retained when $l_{\max}
\ll B$.  This error is much smaller than $O(\epsilon)$, which is expected from 
a naive analysis based on the discarded weight $\epsilon = 1 - W_t$.  Here,
there is subtle worry that it may be that truncation works especially well for 
our chosen hopping Hamiltonian \eqref{eqn:hopping} is so local.  A Hamiltonian 
with longer range hopping would have the same Fermi sea and hence the same 
density matrix, but the truncation errors might be worse.

It would be desirable to also calculate the dispersion relation within the
weight-ranked DM truncation scheme, and compare the results to those obtained
within the operator-based DM truncation scheme.  However, in the latter case 
it is problematic even to define the question, since each retained density 
matrix eigenstate would be a many-particle state.  The new truncated 
Hamiltonian might be conveniently expressed in terms of the pseudo-creation 
operators $\{f_l\}$, but many combinations of occupations would not exist.  

The situation would be somewhat analogous to taking a simple, noninteracting 
hopping Hamiltonian for spinfull fermions, and imposing a Gutzwiller projection 
(no doubly occupied sites).  In effect, the projection made a noninteracting 
model into an interacting one.  Similarly a weight-ranked truncation must 
introduce spurious interactions.  Hence, even if a system containing several 
blocks were to be exactly diagonalized (using the truncated basis) we could 
not immediately identify the elementary excitations.  One would need, for 
example, to numerically compute a spectral function $S(q, \omega)$, where 
$(\hbar q, \hbar\omega)$ are momentum and energy, and then locate peaks as a 
function of $\hbar\omega$.  On the other hand, a system which is truncated
using the operator-based truncation scheme can still be represented by a set
of creation and annihilation operators.

\subsubsection{Inadequacy of Plane-Wave Truncation Scheme}

Following this, we argued, based on the real-space structure of the
one-particle density matrix eigenfunctions shown in 
Section~\ref{sect:wavefunctionscaling}, that an operator-based 
truncation scheme can also be defined naturally using the basis formed by 
single-particle plane wave (PW) states on the block of $B$ sites.  
The dispersion 
relation was calculated within this operator-based PW truncation scheme, and 
compared to that calculated within the operator-based DM truncation scheme.  
We find that, other than getting the dispersion exactly right at the zone 
center of the reduced Brillouin zone, the operator-based PW truncation scheme 
is generally inferior to the operator-based DM truncation.  Instead of 
decaying exponentially as $l_{\max}$, the error in the dispersion relation 
calculated within the operator-based PW truncation scheme decays as a power 
law $l_{\max}^{-1}$, which means that more single-particle basis states need 
to be retained in the operator-based PW truncation scheme as compared to the 
operator-based DM truncation scheme.

\subsection{Towards Interacting Systems in Dimensions \boldmath$d \geq 2$}

It is not much harder in principle to analyze the proposed operator-based
density matrix truncation scheme for \emph{noninteracting} fermions in 
two dimensions.  However, it will be harder to
understand the scaling since we cannot simply rank the eigenvalues.  One
dimension was special because we know that each successive state has one more
node than the previous one.  A further very important difference is that in
$d = 1$ there are just two Fermi points, whereas in $d \geq 2$ there is a Fermi
surface.  Thus, whereas in $d = 1$ the density matrix eigenstates near the
pseudo-Fermi level are related to the energy eigenstates at $\pm k_F$, in
$d \geq 2$ these eigenstates will correspond to mixtures of wavevectors from
every piece of the Fermi surface.

Obviously, noninteracting systems do not require numerical studes, so we must
clarify how our results are relevant to the problem of interacting systems.
Firstly, many (gapless) systems of interest are in a phase --- Fermi liquid,
$d$-wave superconductor --- which are noninteracting in the low-energy,
long-wavelength limit.  When applied to a Fermi liquid system, we expect 
(to the extent that the truncation has separated out the low energy modes) 
that any iterative renormalization scheme will converge on the noninteracting 
limit, and it must behave properly in that limit to have even the hope of 
success.  Hence, for a density matrix-based scheme, the first order of 
business is to study the density matrix for a free fermion ground state (as 
in this paper) or for a BCS state,\cite{chung01} i.e.~that the density matrix 
will actually have the simple operator-based form, and hence its many-particle 
eigenstates of the density matrix really are built from the single-particle 
eigenstates.

For an interacting system, the truncated basis should of course be
constructed using the many-body density matrix for that system (not the
noninteracting system).  We expect that the operators generating this
many-particle truncated basis will \emph{not} just be those that create
the 1-particle density matrix eigenstates (they were in the noninteracting
case studied in this paper).  More thought will be needed as to discover
the best recipe to optimize the truncation rule so as to balance the needs
of sectors with different particle numbers in a strongly interacting system.
The simple algebraic structure of the noninteracting density matrix
eigenstates, considered in this paper, has motivated investigations (in
progress) of the relationships among the many-particle density matrix
eigenstates for an interacting system.

A separate reason why our results for the noninteracting fermions are
relevant to the study of interacting systems is that the scaling behaviour
of the noninteracting density matrix should be a good guide to that of
interacting systems, although details may differ.  This is in the same sense 
that mean-field theory is a good guide to the overall pattern of critical 
phenomena.

However, other interacting models of interest sit at quantum critical points
that are not described by quasiparticle interactions, or possess 
fractionalized excitations.  Since we do not presently understand the proper
way to write their wavefunctions in terms of a spatially blocked basis, nor
the proper renormalization step to capture the interblock correlations in
the fractionalized systems, we do not know if a plain block density matrix
gives the proper basis for truncation of the states.  Furthermore, in the
absence of an analytic construction of the block density matrix, for
example, for Laughlin's quantum Hall wavefunction or the one-dimensional
Su-Schrieffer state, we cannot proceed to scaling studies of large blocks
like those found in the present paper, but numerical studies of such density
matrices might be an illuminating subject for future research.

\begin{acknowledgments}
This research is supported by NSF grant DMR-9981744, and made use of
the computing facility of the Cornell Center for Materials Research (CCMR) 
with support from the National Science Foundation Materials Research 
Science and Engineering Centers (MRSEC) program (DMR-0079992).
\end{acknowledgments}

\end{document}